%% file: template.tex
\documentclass{article}

\usepackage{arxiv}

\usepackage[utf8]{inputenc} 
\usepackage[T1]{fontenc}    
\usepackage{hyperref}       
\usepackage{url}            
\usepackage{booktabs}       
\usepackage{amsfonts}       
\usepackage{nicefrac}       
\usepackage{microtype}      
\usepackage{lipsum}		
\usepackage{graphicx}
\usepackage{natbib}
\usepackage{doi}

\title{UX Research on Conversational Human-AI Interaction: A Literature Review of the ACM Digital Library}

\author{\href{https://orcid.org/0000-0003-0368-0032}
 Qingxiao Zheng \\
  School of Information Sciences\\
  University of Illinois at Urbana-Champaign\\
  \texttt{qzheng14@illinois.edu} \\
   \And
 Yiliu Tang \\
  School of Informatics\\
  University of Illinois at Urbana-Champaign\\
  \texttt{yiliut2@illinois.edu} \\
   \And
 Yiren Liu \\
  School of Informatics\\
  University of Illinois at Urbana-Champaign\\
  \texttt{yirenl2@illinois.edu} \\
   \And
 Weizi Liu \\
  College of Media\\
  University of Illinois at Urbana-Champaign\\
  \texttt{weizil2@illinois.edu} \\
  \And
 Yun Huang \\
  School of Information Sciences\\
  University of Illinois at Urbana-Champaign\\
  \texttt{yunhuang@illinois.edu} \\
}

\date{-}

\renewcommand{\shorttitle}{\textit{UX Research:} Conversational Human-AI Interaction}

\begin{document}
\maketitle

\begin{abstract}
Early conversational agents (CAs) focused on \textit{dyadic} human-AI interaction between humans and the CAs, followed by the increasing popularity of \textit{polyadic} human-AI interaction, in which CAs are designed to mediate human-human interactions. CAs for \textit{polyadic} interactions are unique because they encompass hybrid social interactions, i.e., human-CA, human-to-human, and human-to-group behaviors. However, research on \textit{polyadic} CAs is scattered across different fields, making it challenging to identify, compare, and accumulate existing knowledge. To promote the future design of CA systems, we conducted a literature review of ACM publications and identified a set of works that conducted UX (user experience) research. We qualitatively synthesized the effects of \textit{polyadic} CAs into four aspects of human-human interactions, i.e., communication, engagement, connection, and relationship maintenance. Through a mixed-method analysis of the selected \textit{polyadic} and \textit{dyadic} CA studies, we developed a suite of evaluation measurements on the effects. Our findings show that designing with social boundaries, such as privacy, disclosure, and identification, is crucial for ethical \textit{polyadic} CAs. Future research should also advance usability testing methods and trust-building guidelines for conversational AI.

\end{abstract}

\keywords{Conversational Agent \and Chatbot \and Conversational AI \and UX Research \and Literature Review}

\input{Body}

\clearpage

\bibliographystyle{unsrtnat}
\bibliography{references}  






\end{document}

%% file: Body.tex
\section{Introduction}

There is a rapidly growing body of literature on conversational agents or chatbots~\citep{adamopoulou2020overview}. As promising Artificial intelligence (AI) technologies,  conversational agents are defined as "software that accepts natural language as input and generates natural language as output, engaging in a conversation with the user"~\citep{griol2013automatic}; chatbots, meanwhile, are computer programs designed to simulate conversation with human users via text \citep{adamopoulou2020overview, adamopoulou2020chatbots}. As these two terms are often perceived as interchangeable~\citep{rapp2021human,mctear2020conversational}, in the remainder of this paper, we refer to both conversational agents and chatbots as CAs. Scholars have shown that these machines are able to compensate for human shortcomings or exceed human capacities~\citep{fox2021relationship, guzman2020artificial, whittaker2018ai}. However, prior works focus on designing and evaluating \textit{dyadic} human-AI interaction, which involve only one-to-one interactions between humans and their CAs \citep{bickmore2005acceptance, schulman2009persuading, xu2017new, kopp2005conversational, anabuki2000welbo}; whereas more recent works start tapping into \textit{polyadic} human-AI interactions that also support human-human interactions~\citep{kim2021moderator, wang2021towards, kim2020bot, toxtli2018understanding, benke2020chatbot}.

Even though there are extensive literature reviews on CAs, e.g., \citep{seering2019beyond, chaves2021should, de2020intelligent, montenegro2019survey, laranjo2018conversational}, they do not address how \textit{polyadic} CAs are designed and evaluated, nor present the effects of using \textit{polyadic} CAs on handling the challenges of human-human interaction~\citep{hohenstein2020ai}.  In this paper, we overview  UX (user experience) research on \textit{polyadic} CAs that 1) interact with more than one user in the same conversation and 2) engage in bidirectional conversations between all parties (human-AI and human-human). These \textit{polyadic} CAs encompass a wide variety of complexities that \textit{dyadic} Human-AI may not encounter, including multi-party interactions, social roles taking, group hierarchy, or social tension~\citep{van1997discourse}. To evaluate the effects of CAs' support in human-human interactions, we need to examine both human-CA behaviors and human-to-human behaviors, as well as human-to-group behaviors potentially. Given the unique challenges of the design space, little is known about how \textit{polyadic} CAs should be designed to address these different aspects of social interaction and how they can positively influence the overall user experience~\citep{seering2019beyond, hohenstein2020ai}.

To fill the void, we conducted a literature review using a mixed-method approach, mapping out what exists in \textit{polyadic} CAs research. Specifically, we screened 1,302 ACM papers and identified 36 papers that designed \textit{polyadic} CAs and 135 that designed \textit{dyadic} CAs, which included user evaluation results. We investigated what fundamental human-human interaction challenges are addressed by these works, what effects on human-human interaction are evaluated, and what issues are overlooked when designing these CAs.

The contributions of this work to HCI and social computing are manifold. First, we summarize the fundamental challenges (i.e., consensus reaching, uneven participation, lack of emotional awareness, etc.) of human-human interaction that \textit{polyadic} CAs are designed to tackle and their promising results. Second, we synthesize the current practices (e.g., design originality, relationship types, social scale, evaluation method, etc.) and areas that fall short of empirical studies. Third, we point out the issues that are overlooked by the researchers and practitioners in these works and propose to design CAs with \textit{boundary awareness} for supporting human-human interaction via conversational AI. Fourth, we present the differences in research interests and evaluation metrics between \textit{dyadic} and \textit{polyadic} CA studies and identify existing gaps and potential directions. Further, we envision several research opportunities on conversational human-AI interaction concerning the theoretical groundings, relational dynamics, functional dimensions, and metaphysical implications.

\section{Related Work}

In this section, we first review briefly the history and the research landscape of CAs (Conversational Agents or Chatbots), and then point out the gaps in the existing literature.

\subsection{Landscape of CAs}
The current landscape of CAs is complex. The history of conversational interfaces could be traced back to the 1960s when text-based dialogue systems were designed to answer questions and for simulated casual conversation~\citep{mctear2016conversational}. Since then, various terms have been used, and terminologies have become overlapping~\citep{rapp2021human}. Historically, several distinctions have emerged in research around the terms, such as text-based and spoken dialogue systems, voice user interfaces, chatbots, embodied conversational agents, and social robots and situated agents~\citep{mctear2016conversational}. In recent years, with the commercialization of CAs, more terms emerged. For example, the website chatbots.org references over 1,300 chatbots across nine categories~\citep{casas2020trends} and provides 161 conversational AI synonyms \citep{mctear2020conversational}. 

Similarly to the terminologies, the typologies of CAs are also not univocally categorized~\citep{rapp2021human}. In the past decade, researchers classified CAs based on multiple criteria, such as interaction modality (text-based, voice-based, mixed)~\citep{hussain2019survey, bittner2019bot, mygland2021affordances, laranjo2018conversational}; 
scale of social engagement (one-to-one, broadcasting, and community-based)~\citep{seering2019beyond}; knowledge domain (open domain or closed domain)~\citep{hussain2019survey, bittner2019bot}; goals (task-oriented or non-task oriented)~\citep{hussain2019survey, gao2018neural}; duration and locus of control~\citep{folstad2018different}; embodiment~\citep{cassell2000embodied}; design approach (rule-based, retrieval based, and generative based)~\citep{hussain2019survey}; platform (mobile, laptop,  web browser, SMS, cells, multimodal platforms)~\citep{klopfenstein2017rise, laranjo2018conversational}, and application domains~\citep{de2020intelligent, ruan2020supporting}.

In this paper, we include the aforementioned terminologies as our study objects, given that the CAs' major functionalities are conversational-based. For example, embodied conversational agents (ECAs) are agents simulating humans' face-to-face conversations and can use their body language when talking to users~\citep{cassell2000embodied}. Four properties distinguish ECAs from CAs: ECAs can recognize and respond to both verbal and nonverbal user inputs;  deliver both verbal and nonverbal outputs; deal with conversational functions; and give signals that help the conversations~\citep{cassell2000embodied}. This work includes the ECA papers that conduct UX research on conversational interaction.

\subsection{Gaps in Existing Literature Reviews of CAs}
Most of the existing empirical studies focus on designing \textit{dyadic} CAs~\citep{skjuve2019help,adam2020ai, cranshaw2017calendar, lee2019caring, santos2020therapist}, without  addressing the unique design challenges of \textit{polyadic} CAs. Thus, existing literature reviews address CAs' properties mainly in the context of \textit{dyadic} conversationalists and the effect of interaction between the Human and CAs~\citep{seering2019beyond}. 
They look at how CAs can increase user engagement, enhance user experience, or enrich the relationship between humans and CAs. For example, one work mapped relevant themes in text-based CAs to understand user experiences, perceptions, and acceptances towards CAs~\citep{rapp2021human}; and one categorized the characteristics of human-CA interactions into conversational, social, and personification characteristics~\citep{chaves2020should}.  Others are more specific, such as studying CAs' emotional intelligence~\citep{pamungkas2019emotionally}, personalization~\citep{kocaballi2019personalization}, trust-building~\citep{rheu2021systematic}, and human-likeness~\citep{van2020human}.

Recently, an increasing number of CAs are designed to support \textit{polyadic} human-AI interaction, where human-human communication is supported by CAs. While \textit{dyadic} CAs are commonly used to communicate with individuals as personal assistants~\citep{sciuto2018hey, modi2004cmradar}, customer service agents~\citep{lundkvist2004customer, shimazu2002expertclerk}, and personal healthcare partners~\citep{bickmore2010response, bickmore2009taking, grolleman2006break}, some emotional intelligent CAs are designed to help team members gain awareness of other group members' emotional changes, report the overall sentiment of each group discussion, and maintain positive emotions during collaborations~\citep{benke2020chatbot, peng2019gremobot}. Such \textit{polyadic} CAs address unique social challenges that the \textit{dyadic} CAs were not designed to meet, e.g., in the context of a group or a community. However, little is known about the empirical use and the effect of \textit{polyadic} CAs on different scenarios of human-human interaction~\citep{hohenstein2020ai,seering2019beyond,Hohenstei2018AI,chynal2018human}. 

Also, researchers and practitioners have increasingly noted a lack of understanding in achieving quality UX for CAs. Although the popularity of CA applications exists in multiple domains, studies repeatedly reported CAs causing both pragmatic issues, in which  chatbots failed to understand or to help users achieve their intended goals~\citep{folstad2021future, folstad2017chatbots, lee2021exploring}, and issues in which CAs failed to engage users over time~\citep{folstad2017chatbots, lee2021exploring}. Thus, we set the survey scope to include papers that conducted user evaluations. 

Our review of the CAs designed for \textit{polyadic} Human-AI interaction can help researchers and practitioners identify common practices, research gaps, lessons learned, and promising areas for future advancements. Specifically, we are interested in addressing the following research questions:

\textbf{RQ1}: \textit{What fundamental challenges of human-human interaction are addressed by these CAs? }

\textbf{RQ2}: \textit{What are the research interests in \textit{dyadic} and \textit{polyadic} CAs? } 

\textbf{RQ3}: \textit{What are the practices of designing the CAs for \textit{polyadic} human-AI interaction? }

\textbf{RQ4}: \textit{What are the effects of using the proposed CAs on human-human interactions? }

\textbf{RQ5:} \textit{What evaluation metrics have been used in understanding user experience?}

\textbf{RQ6:} \textit{What issues are overlooked by scholars when designing \textit{polyadic} CAs?}

\section{Method}
\begin{figure*}[htbp] 
    \centering           
    \includegraphics*[width=0.95\textwidth]{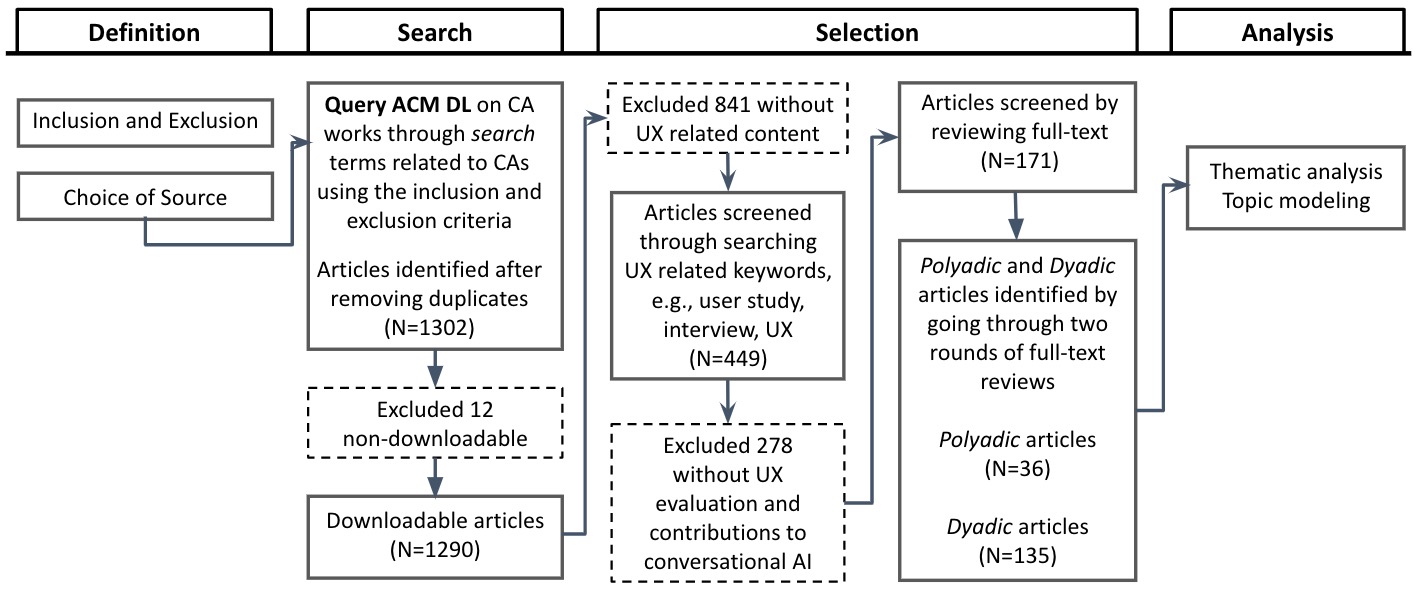}
  \caption{A Workflow Diagram of the Literature Review Process}
  \label{workflow}
\end{figure*}

We used a literature review method that has been widely applied by prior works in HCI, e.g., ~\citep{rapp2021human, nunes2015self, mencarini2019designing}, which has four stages: 1) Define: proposing the inclusion and exclusion criteria and identifying the appropriate data sources; 2) Search: developing specific query and collecting the papers through the data source; 3) Select: checking the search results against the inclusion and exclusion criteria and identifying the final papers for both \textit{dyadic} and \textit{polyadic} works; and 4) Analyze: examining the selected papers by applying a mixed-method approach. We illustrate the four steps in Figure~\ref{workflow} and present the details of each stage below.  

\subsection{Definition}\label{scope}
In this section, we present how we defined the selection criteria and how we identified the appropriate data source.

\subsubsection{Inclusion and Exclusion}

A series of criteria were developed by researchers through multiple rounds of discussions. Criteria were selected to include works that were the most representative of the scope of our study(i.e., UX research on conversational human-AI interaction) and to filter irrelevant works. We employed the following inclusion and exclusion criteria. 

For \textit{dyadic} papers, we defined the inclusion criteria as follows: 1) selected articles need to study CAs that interact with only one human user in a session; 2) the CA interaction in the article is bidirectional; 3) users of the CA are aware of the existence of the CA; 4) the articles are research papers with user studies; 5) the major design feature is conversation-based, e.g., excluded sensory-based CA \citep{bohus2011multiparty}; 6) the articles are written in English; 7) the articles are included in the selected database. Articles that only assessed CA task performance without meaningfully exploring the interaction experience of Human-CA as a conversational technology were excluded. 

For \textit{polyadic} papers, the inclusion criteria shared 2)-7) requirements of selecting \textit{dyadic} papers', except that the selected articles need to study CAs that interact with greater than one human user, instead of with only one human user. Exclusion criteria were defined as: 1) articles that only assessed CA task performance without meaningfully exploring the interaction experience of Human-CA or Human-Human as a conversational technology; and 2) the CAs in the articles interacted with multiple users, but there were no human-human interactions.

\subsubsection{Source}
To identify the data source, we first randomly retrieved 200 papers from five databases: ACM Digital Library, IEEE, Web of Science, Scopus, and Science direct. After exploring the initial search results, we chose Association for Computing Machinery Digital Library (ACM DL) as the final source for our literature review. Two co-authors reviewed 40 papers from each source, using the inclusion and exclusion criteria among all five databases. The qualification rates were ACM DL (23.5\%), IEEE (12.5\%), Web of Science (14.6\%), Scopus (17\%), and Science direct (4.9\%). Specifically, 52\% IEEE papers were technical papers without user evaluations; 59\% Web of Science papers contained no CA designs. Due to the low qualification rates of other sources, we ultimately decided to choose ACM DL as the data source for this literature review. This decision was also made by considering that the ACM DL features a wide selection of reliable HCI works, and has been used as a solo source for literature review works, e.g.,~\citep{ralph2020acm, wainer2009empirical}.

\subsection{Search}
The search query is composed of two parts. The first part of the search covers synonyms of conversational agents, and the second part specifies terms that may be relevant for identifying the \textit{polyadic} papers. The list of search terms was inspired by several CA research reviews \citep{laranjo2018conversational, de2020intelligent, rapp2021human, ter2020design} and was developed through several iterations and refinement by the research team, in line with prior HCI work.

More specifically, the search query included 15 terms, which are “conversational agent”, “conversational AI”, “intelligent assistant”, “intelligent agent”, “chatbot”, “chatterbot", “chatterbox”, “socialbot”, “digital assistant”, “conversational UI”, “conversational interface”, “conversation system”, “conversational system”, “dialogue system“, and “dialog system”. To explore \textit{polyadic} works from the retrieved results, we searched 13 terms, “human-human,” “human human,” “multi-user,” “multi-users,” “multi user,” “multi users,” “multi-party,” “multi-parties,” “multi party,” “multiparty-based”, “multi parties”, "multi model", and  "multi-model."

\subsubsection{Data Preparation}
We searched the papers through two steps, as illustrated in Figure~\ref{workflow}. First,  we searched on Mar $3^{rd}$, 2021 by using the query and connectors throughout the ACM DL, which resulted in 1,302 CA papers after removing duplicates. Second, we web-crawled the metadata and PDFs of these papers and built our database. To improve the stability and efficiency of the crawling process, we also saved the paper DOIs into a file and then crawled the metadata and the PDFs of the respective DOIs in sequence.

\subsubsection{Database}
The database consisted of two parts. The first part was a CSV file that comprehensively covers the metadata of the papers. It included the ten columns: paper DOI, title, authors, abstract, publication date, source, publisher, citations, keywords, affiliation of authors. The second part included the PDFs of the papers, as well as their corresponding text files. We used the Fitz module within the PyMuPDF project~\citep{jorj2016} to transform the PDF to text, as the tool has been applied in a number of works \citep{park2019scientific, yang2019pipelines, park2020figure, chang2021supervised}.

\subsection{Selection}
Once having the database, we screened the paper records to check the \textit{dyadic} and \textit{polyadic} CA works that meet our criteria through two steps,  as illustrated in Figure~\ref{workflow}. We looked for papers that potentially had user studies by searching through the full texts of all papers with the keywords "user study", "user studies", "interview", "interviews" and "user experience", which led to 449 papers. To initiate the set of \textit{polyadic} papers,  we also searched through all the papers using the 13 key terms, e.g., “human-human,” “human human,” “multi-user," which resulted in an initial set of \textit{polyadic} with 27 papers. We then removed duplicates between the two screening results. There were six (1.3\%) duplicates between the two stages, resulting in 470 unique full texts. 
We then further reviewed the full texts against the identified inclusion and exclusion criteria for \textit{polyadic} and \textit{dyadic} papers. If ambiguities persisted about the eligibility of a specific paper, input was sought from the third co-author. 
 The final study collected 36 (21.1\%) \textit{polyadic} papers and 135 (78.9\%) \textit{dyadic} papers.

\subsection{Analysis}
To examine the research landscape of UX research on conversational human-AI interaction and the differences between \textit{polyadic} and \textit{dyadic} works published in ACM DL, we analyzed the selected papers by applying a mixed-method approach.

\subsubsection{Thematic Analysis}
Prior literature review works~\citep{rapp2021human, hussain2019survey, bittner2019bot, laranjo2018conversational} have proposed several aspects when examining conversational AI research, e.g., regarding interaction modality~\citep{hussain2019survey}, characteristics of embodiment~\citep{bittner2019bot} , application domains ~\citep{rapp2021human}, evaluation methods~\citep{rapp2021human}, and social scale~\citep{bittner2019bot}. We found that these could be leveraged to code the research practices  of \textit{polyadic} CAs. However, the prior schemes were not sufficient for us to categorize the unique aspects of human-human interactions supported by \textit{polyadic} CAs, e.g., fundamental challenges addressed, proven effects, and remaining issues. Thus, the authors adopted the grounded theory approach~\citep{wolfswinkel2013using} and coded the attributes of the works using thematic analysis~\citep{braun2006using}. Two co-authors started reviewing the papers and did open coding of all papers independently, and then discussed and compiled their codes together. These codes were added as new attributes to address the proposed RQs. After reaching an agreement regarding the codes, two co-authors  coded the rest papers separately, compared their codes, and discussed possible revisions~\citep{mcdonald2019reliability, pina2018latino}. A final inter-rater reliability of 91\% was achieved and deemed satisfactory~\citep{gwet2014handbook}.

\subsubsection{Topic Modeling} Besides coding the research challenges and issues that are overlooked by scholars in designing \textit{polyadic} CAs, we also applied topic modeling to explore the differences in discussion topics between the two groups of papers. The topic modeling method has been widely applied in prior works \citep{gurcan2021mapping, gurcan2020research, sun2017discovering}. 
This method can be employed as follows. We started text pre-processing by removing punctuation marks, symbols, and stop-words. We then tokenized and lemmatized the text for regularization. 
We used lemmatization to group words with similar semantic meanings but different syntactic forms (such as plurality and tense). 
We experimented with both lemmatization and no lemmatization, and the results with lemmatization exhibited higher interpretability and were more coherent when we sampled the papers to explain the results.  
We then used non-negative matrix factorization (NMF) to conduct topic modeling based on the TF-IDF scores. We built two topic models separately for \textit{dyadic} and \textit{polyadic}  works to explore the major topics discussed in either set of the papers. This provided closer insights into the frequent topics discussed by \textit{dyadic} and \textit{polyadic} CA papers. 
All text analysis steps were completed using Python (NLTK and Scikit-learn libraries). 

\section{Findings}
In this section, we present the findings of the ACM papers, 36 \textit{polyadic} and 135 \textit{dyadic} works, in order of the proposed RQs. Specifically, we show the fundamental challenges~(RQ1 in Section \ref{RQ-challenge}), research interests~(RQ2 in Section \ref{RQ-status}),  practices~(RQ3 in Section \ref{RQ-practice1}), proven effects~(RQ4 in Section \ref{RQ-effects}), evaluation metrics~(RQ5 in Section \ref{RQ-metrics}), and issues~(RQ6 in Section \ref{RQ-issues}).

\subsection{Addressing Fundamental Challenges of Human-Human Interaction (RQ1)}~\label{RQ-challenge}

To answer our first research question, we identified major challenges in human-human interaction addressed in \textit{polyadic} papers. We found that only some papers explicitly stated these challenges as pain points that motivated their designs. Most \textit{challenges} addressed by \textit{polyadic} CAs are in the collaborative learning, work, or group discussion contexts. 

\subsubsection{Inefficient Communication} 
As mentioned in prior research regarding group work and collaborations, several issues in human-human communication affect the efficiency. For example, in online group settings, team communications are often unstructured and less organized, with procrastination and distractions. Reaching consensus can be time-consuming for deliberative discussions~\citep{kim2021moderator,kim2020bot}, and thoughts and discussions can be hard to organize and make sense of ~\citep{zhang2018making} and challenging to control in terms of conversational flow~\citep{bohus2009dialog,bohus2009learning,bohus2010facilitating}. In terms of workflow, task management such as refining, assigning, and tracking can be difficult~\citep{toxtli2018understanding}; heavy workloads related to coordinating schedules among multiple parties can be tedious~\citep{cranshaw2017calendar}; and switching between tools and platforms during collaborations can be burdensome~\citep{avula2018searchbots}. Therefore, \textit{polyadic} CAs are used to provide workflow support and 
discussion support to improve team communication efficiency.

\subsubsection{Lack of Engagement} 
Inactive engagement is a common issue in multi-user interactions in collaborative teams. For example, in peer learning settings, it is challenging to engage students in provocative discussions~\citep{dohsaka2009effects}, to promote productive talk among students, such as explaining to peers and re-voicing others' statements~\citep{dyke2013towards,tegos2015promoting}, and to provide effective learning supports to others, e.g., by exhibiting an agreeable attitude and precipitating tension release~\citep{ai2010exploring, chaudhuri2009engaging, kumar2010socially,kumar2014triggering}. This is also true for collaborative work and online community contexts, in which more inputs~\citep{savage2016botivist} and greater community activity are needed~\citep{seering2018social, seering2020takes, luria2020designing, abokhodair2015dissecting, wang2021towards}. Further issues of engagement in these contexts include uneven participation, in which a balanced amount of contribution is hard to achieve~\citep{kim2020bot, shamekhi2018face}, and user attractions are challenging to capture in casual social settings~\citep{otogi2013finding, otogi2014analysis, zheng2005designing, yuan2005design}.

\subsubsection{Barriers in Relational Maintenance} 
One major challenge in multi-user social settings is maintaining positive relationships, which is crucial to forming a solid team or group. For example, in online collaborative work, there could be a lack of emotional awareness and mutual understanding between team members, as it is hard for them to detect and regulate emotions~\citep{benke2020chatbot,peng2019gremobot,narain2020promoting}.  Moreover, it is always important to grow trust within a team, and the feasibility of using CAs for trust-building~\citep{strohkorb2018ripple} and setting privacy boundaries~\citep{luria2020social} are explored and developed.

\subsubsection{Need for Building Connections} 
In human-human interactions, there is a need for building social connections and identifying common grounds. This is particularly important in the "ice-breaking" and transitioning stages~\citep{low2020gratzelbot}. Also, people may seek similarities and agreement with their communication partners~\citep{nakanishi2003can, isbister2000helper} or need idea supports during chats~\citep{Hohenstei2018AI}. This challenge could be more difficult in cross-cultural conversations. When people from different backgrounds meet for the first time and try to get acquainted with each other, it can be challenging to form impressions of each other free from the influence of cultural stereotypes~\citep{isbister2000helper}. 

\subsection{Research Interests in \textit{Dyadic} and \textit{Polyadic} CAs (RQ2)\label{RQ-status}}
\subsubsection{Research Interest Over Time}

The ACM papers we collected included UX research on  conversational human-AI interaction starting from early 2000, e.g.,  \textit{polyadic} CAs by Isbister et al.~\citep{isbister2000helper} and \textit{dyadic} CAs by Chai et al.~\citep{chai2001conversational}. The works collected were not evenly distributed over the years, e.g., about four UX papers in 2010 and declining to one paper in 2012 for both. However, starting from 2018, there was a surge in the number of selected articles, with the number of \textit{dyadic} papers reaching a peak of 48 in 2020 and the number of \textit{polyadic} papers reaching a peak of nine in the same year.  

\subsubsection{Authors' Affiliations}
There was a total of 596 authors for the 135 \textit{dyadic} papers, averaging 4.4 authors per paper, and 379 (63.6\%) out of the 596 authors were from academia. In comparison, there were a total of 145 authors for the 36 \textit{polyadic} papers, averaging four authors per paper; and 121 (83.4\%) out of the 145 authors were from academia.

\subsubsection{Impactful Works in the Area}
As shown in Table~\ref{dyadiccitations}, our database collected a good body of \textit{dyadic} papers. For example, Medhi et al.~\citep{medhi2011designing} evaluated the usability of a task-oriented CA for novice and low-literacy users and was highly cited, followed by several other usability studies of CAs  \citep{jain2018evaluating, jain2018convey, lau2010conversational} on improving health~\citep{lisetti2013can, bickmore2005acceptance}, answering questions~\citep{liao2018all, lovato2019hey}, and counseling~\citep{schulman2009persuading, smyth2010moses} via CAs. Similarly, a sample of \textit{polyadic} papers received high citations, as  shown in Table~\ref{polyadiccitations}.  For example,  Isbister et al.~\citep{isbister2000helper} designed a virtual agent to support human-human communication in virtual environments. Unlike the \textit{dyadic} works, many selected \textit{polyadic} works conducted UX research in the context of group environments, e.g., virtual meetings~\citep{isbister2000helper, nakanishi2003can}, games~\citep{bohus2010facilitating, bohus2009dialog}, online communities~\citep{savage2016botivist}, and group 
collaboration~\citep{zhang2018making, avula2018searchbots}. 
Our dataset included both \textit{dyadic} and \textit{polyadic} on improving work productivity \citep{kim2019comparing, vtyurina2017exploring, kocielnik2018designing,shamekhi2018face, toxtli2018understanding,cranshaw2017calendar} and promoting education \citep{tanaka2015automated,kumar2010socially, ai2010exploring, dyke2013towards}. 
The average number of citations of a \textit{polyadic} paper in our database was 19 (SD=23) from ACM and 44 (SD=46) from Google Scholar. Five (13.9\%) out of the 36 \textit{polyadic} papers had no citations yet. The average number of citations of a \textit{dyadic} paper in our database was 10 (SD=14) from ACM and 19 (SD=22) from Google Scholar, on December $21^{st}$, 2021. The dataset included works demonstrating impacts at different levels.

\begin{table*}
\caption{A Sample of \textit{Dyadic} ACM Papers that Conducted UX Research.}
\resizebox{\textwidth}{!}{%
\begin{tabular}{llrr}
\hline
\textbf{Year} & \textbf{Authors} & \multicolumn{1}{l}{\textbf{ACM Citations}} & \multicolumn{1}{l}{\textbf{Google Scholar Citations}} \\ \hline
2011  & Medhi, I., Patnaik, S., Brunskill, E., Gautama, S. N. N., Thies, W., \& Toyama, K. \citep{medhi2011designing} & 166 & 275 \\
2018  & Jain, M., Kumar, P., Kota, R., \& Patel, S. N. \citep{jain2018evaluating} & 75 & 187 \\
2013  & Lisetti, C., Amini, R., Yasavur, U., \& Rishe, N. \citep{lisetti2013can} & 85 & 168 \\
2005 & Bickmore, T. W., Caruso, L., \& Clough-Gorr, K. \citep{bickmore2005acceptance} & 62 & 143 \\
2009  & Schulman, D., \& Bickmore, T. \citep{schulman2009persuading} & 50 & 99\\
2018  & Liao, Q. V., Hussain, M. M., Chandar, P., Davis, M., Khazaeni, Y., Crasso, M. P., & 37 & 48 \\ 
&  Wang, D. K., Muller, M., Shami, N. S., \& Geyer, W. \citep{liao2018all}  &  & \\
2019 & Kim, S., Lee, J., \& Gweon, G. \citep{kim2019comparing} & 35 & 84 \\
2017 & Vtyurina, A., Savenkov, D., Agichtein, E., \& Clarke, C. L. A. \citep{vtyurina2017exploring} & 30 & 81 \\
2019 & Lovato, S. B., Piper, A. M., \& Wartella, E. A. \citep{lovato2019hey} & 32 & 65 \\
2018  & Kocielnik, R., Avrahami, D., Marlow, J., Lu, D., \& Hsieh, G. \citep{kocielnik2018designing} & 47 & 46 \\
2010  & Smyth, T. N., Etherton, J., \& Best, M. L. \citep{smyth2010moses} & 37 & 86 \\
2015 & Tanaka, H., Sakti, S., Neubig, G., Toda, T., Negoro, H., Iwasaka, H. \& Nakamura, S. \citep{tanaka2015automated} & 31 & 48 \\
2019 & Zhou, M. X., Mark, G., Li, J., \& Yang, H. \citep{zhou2019trusting} & 30 & 54 \\

2018 & Jain, M., Kota, R., Kumar, P., \& Patel, S. N. \citep{jain2018convey} & 20 & 54 \\
2010 & Lau, T., Cerruti, J.,  Manzato, G., Bengualid, M., Bigham, J. P., \& Nichols, J. \citep{lau2010conversational} & 25 & 44 \\
2005 & Marti, S., \& Schmandt, C. \citep{marti2005physical} & 24 & 40 \\

\end{tabular}%
}
\label{dyadiccitations}
\end{table*}

\begin{table*}
\caption{A Sample of \textit{Polyadic} ACM Papers that Conducted UX Research.}
\resizebox{\textwidth}{!}{%
\begin{tabular}{llrr}
\hline
\textbf{Year}  & \textbf{Authors} & \multicolumn{1}{l}{\textbf{ACM Citations}} & \multicolumn{1}{l}{\textbf{Google Scholar Citations}} \\ \hline
2000  & Isbister, K., Nakanishi, H., Ishida, T., \& Nass, C. \citep{isbister2000helper} & 88 & 207 \\
2010  & Bohus, D., \& Horvitz, E. \citep{bohus2010facilitating} & 81 & 149 \\
2016  & Savage, S., Monroy-Hernandez, A., \& Höllerer, T. \citep{savage2016botivist} & 56 & 113 \\
2009  & Bohus, D., \& Horvitz, E. \citep{bohus2009learning} & 55 & 96 \\
2009  & Bohus, D., \& Horvitz, E. \citep{bohus2009dialog} & 47 & 95 \\
\multicolumn{1}{r}{2018}  & Sebo, S. S., Traeger, M., Jung, M., \& Scassellati, B. \citep{strohkorb2018ripple} & 43 & 76 \\
2018  & Shamekhi, A., Liao, Q. V., Wang D., Bellamy, R. K. E., \& Erickson, T. \citep{shamekhi2018face} & 46 & 73 \\
2018  & Zhang, A. X., \& Cranshaw, J. \citep{zhang2018making} & 39 & 61 \\
2010  & Kumar, R., Ai, H., Beuth, J. L., \& Rosé, C. P. \citep{kumar2010socially} & 27 & 68 \\
2018  & Toxtli, C., Monroy-Hernández, A., \& Cranshaw, J. \citep{toxtli2018understanding} & 35 & 59 \\
2017 & Cranshaw, J., Elwany, E., Newman, T., Kocielnik, R., Yu, B. W., Soni, S., Teevan, J., & 21 & 67 \\
& \& Monroy-Hernández A. \citep{cranshaw2017calendar} & & \\
2003 & Nakanishi, H., Nakazawa, S., Ishida, T., Takanashi, K., \& Isbister, K. \citep{nakanishi2003can} & 9 & 65 \\
2018 & Avula, S., Chadwick, G., Arguello, J., \& Capra, R. \citep{avula2018searchbots} & 20 & 42 \\
2010 & Ai, H., Kumar, R., Nguyen, D., Nagasunder, A., \& Rosé, C. P. \citep{ai2010exploring} & 3 & 56 \\
2013 & Dyke, G., Howley, I., Adamson, D., Kumar, R., \& Rosé, C. P. \citep{dyke2013towards} & 1 & 52 \\
2018 &  Seering, J., Flores, J. P., Savage, S., \& Hammer, J. \citep{seering2018social} & 23 & 29
\end{tabular}%
}
\label{polyadiccitations}
\end{table*}

\subsection{Practices of Designing CAs for \textit{Polyadic} Human-AI Interaction (RQ3)}\label{RQ-practice1}

In the following section, we present aspects that are often shared by \textit{polyadic} and \textit{dyadic} CA works and aspects that are unique to \textit{polyadic} CA works.

\subsubsection{Shared Aspects between \textit{Polyadic} and \textit{Dyadic} CAs} 

We present aspects that \textit{polyadic} and \textit{dyadic} CAs shared, i.e., application domain, modality, agent characteristics, design originality, design method, and evaluation methods below.

\textit{Application domain}. Among the 36 ACM papers we reviewed, eight \textit{polyadic} CAs have been applied to education and collaborative learning,  e.g., \citep{dyke2013towards}; five to online communities, e.g., \citep{savage2016botivist}; three to group discussions, e.g., \citep{kim2021moderator}; seven to work and productivity, e.g., \citep{toxtli2018understanding, avula2018searchbots}), two to virtual meetings, e.g., \citep{isbister2000helper}; three to guiding services, e.g., \citep{zheng2005designing}; two to games, e.g., \citep{rehm2008she}, one to family, i.e., \citep{luria2020social}, and five undefined depending on the major focuses of the papers.

\textit{Modality}. There are 22 polyadic papers applying text-only interactions, e.g., \citep{peng2019gremobot}; nine studies are based on video, e.g., \citep{nakanishi2003can}, only a few are audio only, e.g., \citep{luria2020social}, and hybrid of audio-text, e.g., \citep{isbister2000helper}. These categories are not mutually exclusive,  because some studies implemented multiple designs with different modalities, e.g.,~\citep{shamekhi2018face}.

\textit{Agent characteristics}. We also looked at whether these CAs in the reviewed papers are embodied or not. While 22 papers studied CAs that are not embodied, e.g., \citep{chaudhuri2009engaging}, the rest of them are embodied CAs, e.g., \citep{isbister2000helper}, which have figures that can demonstrate certain social characteristics through animated body languages \citep{cassell2000embodied}.

\textit{Design originality}. Most studies present and evaluate original designs, e.g.,~\citep{chaudhuri2009engaging} while three test or adopt existing systems. 
One paper uses existing systems (i.e., Google Allo) ~\citep{Hohenstei2018AI} and two papers examine existing bots on the Twitter platform~\citep{seering2018social, abokhodair2015dissecting}.

\textit{Design methods}. Among those papers which specified their design methods, five of them are framework-based, which means that the designed features are proof-of-concept designs guided by prior frameworks, models, or theories~\citep{dyke2013towards, tegos2015promoting, kim2021moderator, wang2021towards, nakanishi2003can}. Meanwhile, most of the designed features are proposed by researchers without leveraging  prior designs or catering to specific user-needs, e.g.,~\citep{toxtli2018understanding, dohsaka2009effects, seering2020takes}. Two CAs adopted participatory design methods, including need-finding interviews, e.g.,~\citep{kim2020bot, zhang2018making} and two ran ideation workshops, e.g., ~\citep{benke2020chatbot, luria2020social}.

\textit{Evaluation methods}. The most common evaluation method in the reviewed studies is experimental, with 16 papers using it. Other methods include six using surveys, six using interviews, five using field studies, three using Wizard of Oz, two using interaction log analysis, two using observation and one using focus groups. These categories are not mutually exclusive since some studies employed multiple methods in their evaluations.

\subsubsection{Unique Aspects with the  \textit{Polyadic} CAs} Below, we present aspects that involve multi-user interaction, unique to \textit{polyadic} CAs, i.e., relationship type and social scale. Moreover, we report related theories and frameworks used in prior \textit{polyadic} works.

\textit{Relationship types}. The specific relationships highly depend on the contexts of the interactions, which include eight papers designing \textit{polyadic} CAs as co-learners, e.g., \citep{chaudhuri2009engaging}, four as co-workers, e.g.,~\citep{cranshaw2017calendar}, seven as collaborators/discussants, e.g., collaborators on Slack~\citep{avula2018searchbots}, three as people meeting for the first time, e.g.,~\citep{isbister2000helper}, four as online community members, e.g., \citep{savage2016botivist}, three as public visitors and guests, e.g., \citep{zheng2005designing}, six as co-players in a game, e.g., \citep{rehm2008she}, one as SMS contacts, e.g.,~\citep{Hohenstei2018AI}, and one as family members, e.g.,~\citep{luria2020social}. Since \textit{polyadic} CAs are designed for multiple users in contrast to \textit{dyadic} CAs for one-on-one interactions, the relationship types between users are a unique dimension for us to examine \textit{polyadic} human-AI interactions.

\textit{Social scale}. The social scale of the human users is another unique dimension of \textit{polyadic} human-AI interactions since the designs examined in these papers are for two-individuals or multi-users (more than three users). There are 18 papers on multi-users. Combining the relationship types, we can see that the multi-users scale ranges from smaller groups (e.g., three to five co-learners, \citep{dyke2013towards}), to medium sized groups (e.g., ten  discussants~\citep{kumar2010socially}), and to larger groups (e.g., online community members on Twitter~\citep{savage2016botivist}).

\textit{Social theories and related theoretical frameworks}. We also collected the theories that motivated the designs of the CAs, i.e., presenting theory-inspired or framework-based designs; other papers are less theory-driven. 
Among them, self-extension theory has been used to discuss users' sense of ownership of a community-owned CA \citep{luria2020designing}; the Computers Are Social Actors (CASA) paradigm has been used to compare the patterns of human-CA interactions and human-human interactions \citep{rehm2008she}; message framing theory has been applied to design the bot messages in implementations and evaluations \citep{savage2016botivist}; balance theory has been leveraged to explore the sense of "balance" in the social dynamics in group interaction mediated by CAs \citep{nakanishi2003can}; and the Structural Role Theory \citep{biddle1986recent} has been used to identify the roles that the bots play on Twitch \citep{seering2018social}. 

Other papers build on theoretical frameworks, such as: proposing a harmony model of CA mediation using Benne's categorization of functional roles in small group communication~\citep{benne1948functional}; positing Mutual Theory of Mind as the framework to design for natural long-term human-CA interactions~\citep{wang2021towards}; and applying Input-Process-Output models~\citep{kozlowski2006enhancing} to explain the process of emotional management in teams~\citep{benke2020chatbot}. A few  papers discuss important concepts in multi-user interactions. For example, vulnerability and trust are discussed in teamwork settings to support a design of robot that is intended to build trust in teams; and Barsade's Ripple Effect \citep{barsade2002ripple} is introduced to support hypotheses of a machine agent's positive influence on other individuals in a team just as effectively as a human member \citep{strohkorb2018ripple}. However, the theories and theoretical frameworks used in the limited number of existing publications are rather scattered and disorganized.

\subsection{Proven Effects of \textit{Polyadic} CAs  on Human-Human Interaction (RQ4) \label{RQ-effects}}

The reviewed articles reported two major sets of effects of \textit{polyadic} CAs: the effects on \textit{dyadic} interaction, emphasizing the effects on one-on-one interaction between human users and the CAs; and the effects on \textit{polyadic} interaction, referring to the impact of the CAs designed to mediate human-human interaction. The effects of the \textit{polyadic} CAs on \textit{polyadic}  human-AI interaction include improving the individual users' learning effectiveness \citep{tegos2015promoting, chaudhuri2009engaging}, perceived self-efficacy \citep{tegos2015promoting}, and satisfaction \citep{dohsaka2009effects}. Another large portion of this set of effects involves users' positive or negative attitudes, perceptions, and acceptance toward the CAs \citep{ai2010exploring, wang2021towards, nakanishi2003can, rehm2008she}, and interactions and engagement patterns between users and the CAs depending on specific design features \citep{ai2010exploring, strohkorb2018ripple}. Most studies laid primary focus on \textit{dyadic} human-AI interaction effects, while the evaluation on group effects is relatively insufficient. Overall, the effects reported in the reviewed papers show opportunities of using \textit{polyadic}  CAs to improve human social experience.

\subsubsection{Communication Efficiency}
In collaborations, studies have supported that \textit{polyadic} CAs can help with consensus-reaching, aid communication comprehension, enhance task management, and save the time and energy of human collaborators from tedious work, which can improve the group work efficiency and overall collaborative experience.
Specifically, for consensus-reaching, a moderator CA helps to align group consensus and individual opinions, contributing to reaching agreements~\citep{kim2021moderator, kim2020bot}. For better comprehension of group communication, a CA can tag and summarize the group chats to help users make sense of the conversation better~\citep{zhang2018making}. Moreover, a tour guide CA in museums can mediate visitors' interactions by prompting topics, offering information, and concluding discussions~\citep{otogi2013finding, otogi2014analysis}. 
In terms of management and coordination, a scheduling CA can coordinate schedules of team members fast and efficiently, setting human personal assistants free~\citep{cranshaw2017calendar}. 
For burdensome tasks, a searchbot is designed for its human collaborators to save time when switching between tools and searching independently by offloading these tasks to the CA.

\subsubsection{Group Engagement}
The papers we reviewed have provided evidence that \textit{polyadic} CAs may benefit group dynamics through encouraging engagement and balancing uneven participation. The effects of encouraging engagement are most salient for collaborations in education and online communities. In multiple studies in collaborative learning, the intervention of a CA that can ask questions and prompt students to show their thought processes, stimulating pedagogically beneficial conversations in the learning groups~\citep{dyke2013towards, tegos2015promoting}. In online communities such as Twitter, bots designed to call people to action to specific social activities engage users to make relevant contributions to the discussion and lead to their interactions  regarding future collaborations~\citep{savage2016botivist}. 

Also, \textit{Polyadic} CAs as facilitators in group discussions not only encourage the amount of engagement, but also improve the distribution of the engagement by nudging members to participate evenly~\citep{shamekhi2018face}. Several CA designs can balance discussion by inviting less engaged learners to join the conversation and downplaying the “talkative” learners from over-controlling conversations~\citep{chaudhuri2009engaging}. The BabyBot on Twitch can grow up and learn from human users, which engages community members as a whole and opens opportunities for newcomers to participate in interactions as a way to welcome them on board~\citep{luria2020designing, seering2020takes}. 

CAs are also designed to moderate multi-user engagement in open and dynamic environments~\citep{bohus2009dialog,bohus2009learning,bohus2010facilitating}, e.g., the hotel reception, games, and organizational systems, by identifying partakers and bystanders, and facilitating engagement decisions of when and whom to engage. Furthermore, more intensive and even engagement brings more diverse content. The moderator CA for discussions can therefore help with generating diverse and deliberative opinions~\citep{kim2021moderator, kim2020bot}.

\subsubsection{Relationship Maintenance}
Regarding social and relational aspects in groups, \textit{polyadic} CAs show potential in regulating emotions and relationships to maintain harmonized group dynamics. Prior work presented prototype designs of CAs to offset a shortcoming of the text-based communication in teams, that is, the insufficient ability of understanding and managing emotions of the team members~\citep{benke2020chatbot, kumar2014triggering}. By monitoring the sentiment of the conversations and providing suggestions, CAs can improve emotion regulation and compromise facilitation \citep{benke2020chatbot}. Similarly, \textit{polyadic} CA can analyze group behaviors to reinforce positivity by preventing the use of negative words~\citep{peng2019gremobot}. 
Moreover, designing an agent that can make vulnerable statements in collaboration generates a \textit{Ripple Effect}, which notes that human teammates with an agent that expresses vulnerability are more likely to engage in trust-related behaviors, including explaining failures to the group, consoling team members who made mistakes, and laughing together. These actions reduce tension in the team and enhance a positive and trusting atmosphere in groups~\citep{strohkorb2018ripple}. 

\subsubsection{Building Connections}
In a two-individuals interaction, the existence of a CA is crucial to the balance and solidarity of the triad. Studies have shown that a CA's agreement, disagreement, or bias toward one or both side(s) of the two individuals can significantly change the mutual perceptions of the interaction partners \citep{ai2010exploring, nakanishi2003can}. Also, when a CA agrees or disagrees with both the human pair, the interaction partners feel more similar and attractive to each other \citep{nakanishi2003can}. In contrast, a CA tutor showing a bias towards one perspective or another will polarize the learning pair \citep{ai2010exploring}. Thus, \textit{polyadic} CAs 
demonstrate potential to build solidarity between interaction pairs. 

\textit{Polyadic} CAs may also be used to establish new connections between pairs that meet for the first time. i.e., a helper agent that is built to form social connections between people in cross-cultural conversations~\citep{isbister2000helper}. The evaluation showed the positive effects of the agent designed to build common ground. Also, a CA was designed as a virtual companion in a university setting to help first-year students get on board and build new connections~\citep{low2020gratzelbot}. However, there were also a series of mixed findings of nuanced human users' social perceptions regarding self, partner, and cultural stereotypes, which are worth further exploration.

\subsection{Evaluation Metrics of CAs (RQ5)\label{RQ-metrics}}

We synthesize the metrics of prior studies used to evaluate user experience of conversational agents. These metrics show what categories previous researchers considered in evaluating their designs and what measurements they used. In the following section, we report on the results emerging from the analysis of the 135 \textit{dyadic} papers and the 36 \textit{polyadic} papers included in the final corpus. We categorize them in three main areas: chat log analysis or observation, survey scale, and interview. For definitions of all the metrics, see Appendix~\ref{append-1} and \ref{append-2} for details.

\subsubsection{Log Analysis or Observations} 
For \textit{dyadic} papers, 36 papers evaluated CAs using log analysis or observation methods such as coding user behaviors through watching recorded videos. Table~\ref{tab:chat_analysis}. provides an overview of what metrics were evaluated in the user studies. Six categories emerged: 1) linguistic features, i.e., language patterns that are evident in the language use; 2) prosodic features, i.e., features that appear when we put sounds together in connected speech; 3) dialogue features, i.e., features that are related to conversation structure and chat flow; 4) tasks-related features, i.e., features that measures metrics related to task fulfillment; 5) algorithm features, i.e., features concerning model training and efficiency; and 6) user-related features, i.e., features that are closely related to user perceptions and behaviors.

Linguistic features were evaluated in seven recent papers. To analyze the linguistic features in the chat logs, most of these studies leveraged existing natural language processing toolkits to analyze the text. For example, LIWC~\citep{kawasaki2020assessing} uses 82 dimensions to determine if a text uses positive or negative emotions, self-references, causal words etc. Another way is to count the use of linguistic patterns, such as the use of pronouns and proportion of utterances with terms repeated from previous conversations~\citep{thomas2020expressions}. Similar to linguistic features, prosodic features were used to evaluate the language pattern, but especially in speech-based agents, e.g., prior work measured the rate of speech, pitch variation, loudness variation, using OpenSMILE to process the audio signals~\citep{thomas2020expressions}.

Dialogue features were used earlier than linguistic features, and were evaluated in nine papers, measuring the dialogue quality and efficiency~\citep{xiao2020if}, e.g., length, dialogue duration, number of turn taking, response time; and dialogue expressiveness, e.g., percentage of sentences. One of the frameworks being widely adopted in prior works is the PARAdigm for dialogue system evaluation, also known as PARADISE framework, which examines user satisfaction based on measures representing the performance dimensions of task success, dialogue quality, and dialogue efficiency, and has been applied to a wide range of systems, e.g.,~\citep{foster2009comparing}. 

Task-related features were examined in nine papers. This dimension helps researchers to understand how users fulfill the tasks assigned by or facilitated by CAs. These dimensions are useful for task-oriented CAs that were designed to help users execute a task or solve a problem, e.g., customer service. While the task fulfillment rate was continually used as a metric in evaluating tasks, e.g., ~\citep{demberg2011strategy,medhi2011designing,pecune2018field}, some metrics were introduced in recent years, such as dropout rate, e.g., ~\citep{kim2019comparing}, to capture the percentage of all participants who did not complete the task during evaluation.

Eight papers evaluated user perceptions and social behaviors by analyzing the chat log between the CA and the users, using metrics such as self-disclosure~\citep{lee2020designing,lee2021exploring}, intimacy~\citep{sannon2018personification}, and aggressiveness~\citep{bonfert2018if}. Some theoretical frameworks were used, e.g., Barak and Gluck-Ofri's scale, which conceptualized self-disclosure into three dimensions as information, thoughts, and feelings~\citep{lee2020designing}.

For \textit{polyadic} papers we reviewed, 14 studies employed log analysis and were mostly published during the 2010s. Several papers used task-oriented metrics, see Table~\ref{tab:HH}, e.g., task duration, efficiency, effectiveness, and team performance \citep{avula2018searchbots,seering2020takes,luria2020designing, kumar2014triggering}. The evaluation metrics were both quantitative and qualitative. The quantitative metrics counted the frequency of users' input or utterances \citep{tegos2015promoting,avula2018searchbots,dohsaka2009effects,abokhodair2015dissecting,otogi2013finding,otogi2014analysis, narain2020promoting}. The qualitative metrics evaluated discussion quality~\citep{wang2021towards, tegos2015promoting, kim2021moderator}, with an emphasis on consensus reaching~\citep{kim2021moderator}, opinion diversity~\citep{tegos2015promoting}, and linguistic features, e.g.,  verbosity~\citep{wang2021towards}, readability~\citep{wang2021towards}, adaptability~\citep{wang2021towards} and positive and negative sentiment~\citep{wang2021towards,Hohenstei2018AI, nakanishi2003can,narain2020promoting}.

\begin{table*}[!ht]
    \caption{\textit{Polyadic} ACM Papers using Different Metrics for UX Evaluation (Count by Year Since 2005)}
    \centering
    \resizebox{\textwidth}{!}{
\begin{tabular}{l@{\hspace{4pt}}l@{\hspace{4pt}}l@{\hspace{1pt}}r@{\hspace{3pt}}r@{\hspace{3pt}}r@{\hspace{3pt}}r@{\hspace{3pt}}r@{\hspace{3pt}}r@{\hspace{3pt}}r@{\hspace{3pt}}r@{\hspace{3pt}}r@{\hspace{3pt}}r@{\hspace{3pt}}r@{\hspace{3pt}}r@{\hspace{3pt}}r@{\hspace{3pt}}r@{\hspace{3pt}}r@{\hspace{3pt}}r@{\hspace{3pt}}r}
\toprule
\textbf{Category} & \textbf{Subcategory} & \textbf{Metrics (Defined in Appendix~\ref{append-1})} & \textbf{(20)05} & \textbf{06} & \textbf{07} & \textbf{08} & \textbf{09} & \textbf{10} & \textbf{11} & \textbf{12} & \textbf{13} & \textbf{14} & \textbf{15} & \textbf{16} & \textbf{17} & \textbf{18} & \textbf{19} & \textbf{20} & \textbf{21}\\
\midrule
Chatlog & Socialbility  & discussion quality &  1 &   &    &    &  1 &  1 &    &    &    &    &    &    &    &    &    &   &    \\
&        &  consensus reaching &    &    &    &    &    &  1 &    &    &    &    &    &    &    &    &    &    &    \\
&        & opinion expression &    &    &    &    &    &  1 &    &    &    &    &    &    &    &    &    &    &    \\
            &        & opinion diversity &    &    &    &    &    &  1 &    &    &    &    &    &    &    &    &    &    &    \\
            &        & even participation &    &    &    &    &    &  1 &    &    &    &    &    &    &    &    &    &    &    \\
            &        & linguistic features &    &    &    &    &    &    &    &    &    &    &    &    &    &    &    &  1 &    \\
            &        & message quantity &    &    &    &    &    &  1 &    &    &    &    &    &    &    &   1 &    &  1 &    \\
& Task-related & task completion time &    &    &    &    &    &    &    &    &    &  1 &    &    &    &    &    &    &    \\
            &        & efficiency &    &    &    &    &    &    &    &    &    &    &    &    &    &    &    &  1 &    \\
            &        & effectiveness &    &    &    &    &    &    &    &    &    &    &    &    &    &    &    &  1 &    \\
            &        & satisfaction &    &    &    &    &    &    &    &    &    &    &    &    &    &    &    &  1 &    \\
            &        & team performance &    &    &    &    &    &    &    &    &    &  1  &    &    &    &    &    &  1 &    \\
Survey&Communication- & perceived communication effectiveness &    &    &    &    &    &  1 &    &    &    &    &    &    &    &    &    &    &    \\
 & quality & perceived communication fairness &    &    &    &    &    &  1 &    &    &    &    &    &    &    &    &    &  1 &    \\
& & perceived  communication efficiency &    &    &    &    &    &  1 &    &    &    &    &    &    &    &    &    &    &    \\
            &        & perceived group climate &    &    &    &    &    &    &    &    &    &    &    &    &    &  1 &    &    &    \\
&Socio-emotional &  perception of other group members  &    &    &    &    &    &    &   &    &    &   1  &  1 &    &    &  1  &    &  1  \\
            &        & perception towards the agent &    &    &    &    &    &    &    &    &    & 1  &    &    &    &   &    &    &    \\
            &        & perception about collaboration &    &    &    &    &    &    &    &    &    &    &    &    &    &    &    &    &    \\
            &        & perceived social presence &    &    &    &    &    &    &    &    &    &    &    &  1 &    &  1 &    &    &    \\
            &        & perceived social support &    &    &    &    &    &    &    &    &    &    &    &  1 &    &    &    &    &    \\
            &        & level of annoyance with the intervention &    &    &    &    &    &    &    &    &    &    &    &    &    &    &    &    &    \\
            &        & opinion of oneself/partner/cultural stereotypes &    &    &    &    &    &  1 &    &    &    &    &    &    &    &    &    &    &    \\
            &        & perceived appropriateness of agent's social behavior &    &    &    &    &    &    &    &    &    &    &    &    &    &  1 &    &    &    \\
            &        & users' psychological wellbeing &    &    &    &    &    &    &    &    &    &    &    &    &    &  &    &  1 &    \\
&Traditional & anthropomrophism &    &    &    &    &    &    &    &    &    &    &    &    &    &  1 &    &    &    \\
            &        & perceived emotional competence &    &    &    &    &    &    &    &    &    &    &    &  1 &    &    &    &    &    \\
            &        & unmet expectations &    &    &    &    &    &    &    &    &    &    &    &  1 &    &    &    &    &    \\
            &        & privacy concerns &    &    &    &    &    &    &    &    &    &    &    &  1 &    &    &    &    &    \\
            &        & level of perceived conversation control &    &    &    &    &  1 &    &    &    &    &    &    &    &    &    &    &    &    \\
            &        & perceived satisfaction &    &    &    &    &    &  1 &    &    &    &    &    &    &    &    &    &  1 &    \\
           &         & performance/effectiveness &    &    &    &    &    &    &    &    &    &    &    &    &    &  1 &    &  1 &    \\
           &        & level of engagement &    &    &    &    &    &    &    &    &    &    &    &    &    &    &    &    &    \\
Interview&Socialbility & perceived capability to promote contributions &    &    &    &    &    &    &    &    &    &    &    &    &    &  1 &    &    &    \\
&Issue-relevant & decisions to (not) engage with the CA &    &    &    &    &    &    &    &    &    &    &    &    &    &    &    &  1 &    \\
           &         & overall UX &    &    &    &    &    &    &    &    &    &    &    &    &    &  1  &    &  1 &    \\
            &        & reflections on selves and others &    &    &    &    &    &    &    &    &    &    &    &    &    &    &    &  1 &    \\
           &         & willingness to take actions &    &    &    &    &    &    &    &    &    &    &    &  1 &    &    &    &    &    \\
           &         & direction of improvement &    &    &    &    &    &    &    &    &    &    &    &   &    &  1  &    &  1  &    \\
\bottomrule
\end{tabular}
}

    \label{tab:HH}
\end{table*}

\begin{table*}[!ht]    
\centering
    \caption{\textit{Dyadic} ACM Papers using Chat Analysis for UX Evaluation (Count by Year Since 2005)}
    \resizebox{\textwidth}{!}{
\begin{tabular}{llrrrrrrrrrrrrrrrrr}
\toprule
            \textbf{Category} &                                \textbf{Metrics (Defined in Appendix~\ref{append-2})} &  \textbf{(20)05} & \textbf{06} & \textbf{07} & \textbf{08} & \textbf{09} & \textbf{10} & \textbf{11} & \textbf{12} & \textbf{13} & \textbf{14} & \textbf{15} & \textbf{16} & \textbf{17} & \textbf{18} & \textbf{19} & \textbf{20} & \textbf{21}\\
\midrule
 Linguistic features &         positive/negative words &      &      &      &      &      &      &      &      &   &      &      &      &      &      &     1 &     3 &      \\
 &                   length of utterances &      &      &      &      &      &      &      &      &      &      &      &      &      &      &      &     2 &      \\
 &                      personal pronouns &      &      &      &      &      &      &      &      &      &      &      &      &      &      &      &     2 &      \\
 &                  term-level repetition &      &      &      &      &      &      &      &      &      &      &      &      &      &      &      &     1 &      \\
 &             utterance-level repetition &      &      &      &      &      &      &      &      &      &      &      &      &      &      &      &     1 &      \\
 &                  use of concrete words &      &      &      &      &      &      &      &      &      &      &      &      &      &      &      &     1 &      \\
 &             variation in language used &      &      &      &      &      &      &      &      &      &      &      &      &      &      &      &     1 &      \\
   Prosodic features &       variation in speech-based agents &      &      &      &      &      &      &      &      &      &      &      &      &      &      &      &     1 &      \\
   Dialogue features &              dialogue/response quality &      &      &      &      &     1 &      &      &      &      &      &      &      &      &      &      &     3 &      \\
    &            dialogue efficiency metrics &      &      &     1 &      &     1 &      &     1 &      &      &      &      &      &      &      &      &     2 &      \\
    &                         expressiveness &      &     1 &      &      &      &      &      &      &      &      &      &      &      &      &      &      &      \\
       Tasks-related &  task fulfillment rate/effectiveness &      &      &      &      &     1 &      &     2 &      &      &      &      &      &      &     1 &      &      &      \\
        &                           system usage &      &      &      &      &      &      &      &      &      &      &      &      &      &     3 &      &      &      \\
        & user entry time &      &      &      &      &      &      &     1 &      &      &      &      &      &      &      &      &      &      \\
        &                           dropout rate &      &      &      &      &      &      &      &      &      &      &      &      &      &      &     1 &      &      \\
           Algorithm &             Intent modeling evaluation &      &      &      &      &      &      &      &      &      &      &      &      &      &      &      &     1 &      \\
            &                             error rate &      &      &      &      &      &      &     1 &      &      &      &      &      &      &      &      &      &      \\
                User related &                       self-disclosure &      &      &      &      &      &      &      &      &      &      &      &      &      &     1 &     1 &     1 &     1 \\
                 &                               intimacy &      &      &      &      &      &      &      &      &      &      &      &      &      &     1 &      &      &      \\
                 &                             reflection &      &      &      &      &      &      &      &      &      &      &      &      &      &     1 &      &      &      \\
                 &         impolite/aggressive behaviors &      &      &      &      &      &      &      &      &      &      &      &      &      &     1 &      &      &      \\
                 &                             motivation &      &      &      &      &      &      &      &      &      &      &      &      &      &      &      &     1 &      \\
                 &                       affective states &      &      &      &      &      &      &      &      &      &      &      &      &      &      &      &     1 &      \\
\bottomrule
\end{tabular}
}

    \label{tab:chat_analysis}
\end{table*}

\subsubsection{Survey Scale}
For \textit{dyadic} papers, 93 papers evaluated CAs using survey scales. Table~\ref{tab:survey_scale} provides an overview of what features were evaluated in the user studies. Four categories emerged: 1) conversation related metrics: aspects that help CAs to manage dialogues during interactions; 2) user perception of CA's social features: aspects that reflect user perception of CA following social protocols; 3) user perceived system usability: aspects that capture the quality of a UX when users interact with CAs; and 4) user self-reported experience with CAs.

For conversation related features, some studies leverage widely adopted scales because they are often used for system evaluation. For example, informativeness, or information quality, has been used multiple times in user evaluations. Informativeness refers to quality of the conversational system to communicate truthfully, provide relevant information, and share content clearly and orderly. Prior work on AI-powered CA used Gricean Maxims' information quality metrics~\citep{xiao2020tell} to evaluate this. Other metrics were proposed as modes of examination in recent years, e.g., syntactic readability, self-reported response quality and intensity of sentiments\citep{wambsganss2020conversational}, and instruction quality~\citep{foster2009comparing}.

Multiple features were proposed in evaluating users perceptions of conversational agents' social features, such as perceived agents' common ground~\citep{chai2014collaborative}, sociability, ~\citep{wang2020alexa}, and perceived interruption or annoyance~\citep{schulman2009persuading}). Metrics such as playfulness~\citep{zhou2019trusting, xiao2019should} (enjoyment, interestingness, or funny) and perceived intimacy (perceived interpersonal closeness, or friendliness) have been evaluated using existing scales. For example, the inclusion of others in the self scale and the subjective closeness index to measure level of closeness~\citep{yu2019almost}. With the development of conversational agents, more social features have been introduced in user evaluation, such as perceived anthropomorphism~\citep{s2020capturing}, perceived personality traits~\citep{volkel2020trick}, and perceived naturalness~\citep{shi2020effects}. 

For system usability evaluation, 33 papers evaluate CA usability using traditional usability metrics. For example, UMUX-LITE is the usability metric for user experience to measure evaluating factors such as perceived ease of use, and it is used in papers~\citep{bickmore2005acceptance,biswas2006flexible}. Similarly, NASA-TLX, or task load index, which evaluates mental demand, physical demand, effort, frustration, and future adoption willingness, is used in studies~\citep{jain2018convey,jain2018evaluating}. Moreover, prior work also uses usability questions from previous surveys ~\citep{khadpe2020conceptual}.

There are 45 papers evaluating the conversational agent through collecting metrics regarding user self-reported personal experience responses. Among them, satisfaction, e.g., ~\citep{demberg2011strategy,bickmore2005acceptance,liu2020cold} and perceived trust, e.g.,~\citep{zhou2019trusting, lau2010conversational,hoegen2019end,lee2020designing}, and perceived engagement, e.g.,~\citep{muresan2019chats,cai2021critiquing,shi2020effects, xiao2020if} were evaluated the most in the sample. There are some new features been used to evaluate \textit{dyadic} interactions, such as serendipity~\citep{cai2021critiquing}, social orientation toward CAs~\citep{ashktorab2019resilient}, degree of collaboration~\citep{mitchell2021automated}, and perceived capability to control~\citep{thomas2020expressions}. Thus, evaluation dimensions are becoming increasingly diverse when examining user experience with conversational agents.

\textit{Polyadic} studies which employed surveys mainly evaluated users' perceptions of both the functional and socio-emotional aspects of the interactions, with specific focus on communication quality and interpersonal dynamics. The former includes perceived communication effectiveness~\citep{tegos2015promoting,kim2021moderator}, communication efficiency~\citep{tegos2015promoting,kim2021moderator,luria2020designing}, communication fairness~\citep{kim2021moderator,kim2020bot}, and the quality of collaboration~\citep{avula2018searchbots}. The latter includes perceived group climate~\citep{gulz2011extending, seering2018social}, social presence~\citep{benke2020chatbot,seering2018social}, social support~\citep{benke2020chatbot}, perceptions on the agent~\citep{dohsaka2009effects,seering2018social,seering2020takes,kumar2014triggering}, appropriateness of the agent's intervention~\citep{avula2018searchbots},  as well as impressions about other group members~\citep{gulz2011extending,seering2020takes,kumar2014triggering}. We observe that during the past decade works evaluated social dynamics in the interactions~\citep{gulz2011extending,benke2020chatbot}. These studies with survey methods also used a series of metrics that are commonly used in \textit{dyadic} interactions, which evaluate the ability and competence of CAs~\citep{avula2018searchbots,benke2020chatbot,seering2018social}, task performances~\citep{seering2018social,seering2020takes}, perceived anthropomorphism~\citep{gulz2011extending,seering2018social,shamekhi2018face}, intelligence~\citep{shamekhi2018face}, likeability~\citep{gulz2011extending}, users' satisfaction of the interaction~\citep{kim2021moderator,seering2020takes}, perceived control of the interaction~\citep{nakanishi2003can}, and level of engagement~\citep{avula2018searchbots,seering2018social}.
\begin{table*}[!ht]
    \caption{\textit{Dyadic} ACM Papers using Survey Scales for UX Evaluation (Count by Year Since 2005)}
    \centering
    \resizebox{\textwidth}{!}{
\begin{tabular}{lllllllllllllllllll}
\toprule
\textbf{Category} & \textbf{Metrics (Defined in Appendix A.2)} &  \textbf{(20)05} & \textbf{06} & \textbf{07} & \textbf{08} & \textbf{09} & \textbf{10} & \textbf{11} & \textbf{12} & \textbf{13} & \textbf{14} & \textbf{15} & \textbf{16} & \textbf{17} & \textbf{18} & \textbf{19} & \textbf{20} & \textbf{21}  \\
\midrule
Conversation related & syntactic readability &    &    &    &    &    &    &    &    &    &    &    &    &    &    &    &  1 &    \\
                                 & self-reported response quality &    &    &    &    &    &    &    &    &    &    &    &    &    &    &    &  1 &    \\
                                 & informativeness/information quality &  1 &    &    &    &    &    &    &    &    &    &    &    &    &  1 &    &  1 &    \\
                                 & conversation smoothness &    &    &    &    &    &    &    &    &    &    &    &    &    &    &  1 &    &    \\
                                 & intensity of sentiments &    &    &    &    &    &    &    &    &    &    &    &    &    &    &    &  1 &    \\
                                 & instruction quality &    &    &    &    &  1 &    &    &    &    &    &    &    &    &    &    &    &    \\
Perception towards CA & perceived common ground &    &    &    &    &    &    &    &    &    &  1 &    &    &    &    &    &    &    \\
                                 & opinion of the CA as a partner &    &    &    &    &  1 &    &    &    &    &    &    &    &    &    &    &    &    \\
                                 & liking attitude &  1 &    &    &    &    &    &    &    &    &    &  1 &    &    &  1 &    &    &    \\
                                 & playfulness/enjoyment &  1 &    &    &    &    &    &    &    &  1 &    &    &    &    &  1 &  3 &  3 &  1 \\
                                 & socialbility &    &    &    &    &    &    &    &    &    &    &    &    &    &  1 &    &  1 &    \\
                                 & perceived anthropomorphism &    &    &    &    &    &    &    &    &    &    &    &    &    &    &    &  2 &    \\
                                 & perceived warmth of the AI system &    &    &    &    &    &    &    &    &    &    &    &    &    &    &    &  1 &    \\
                                 & impression &    &    &    &    &    &    &    &    &    &    &    &    &    &    &    &  1 &    \\
                                 & perceived personality traits &    &    &    &    &    &    &    &    &    &    &    &    &    &    &    &  1 &    \\
                                 & perceived persuasiveness &    &    &    &    &  1 &    &    &    &    &    &    &    &    &    &    &    &    \\
                                 & interaption/annoyance &  1 &    &    &    &    &    &    &    &    &    &    &    &    &    &    &    &    \\
                                 & perceived intimacy &  1 &    &    &    &    &    &    &    &    &    &    &    &  1 &  1 &  2 &  2 &  1 \\
                                 & perceived naturalness &    &    &    &    &    &    &    &    &    &    &    &    &    &    &    &  1 &    \\
                                 & perceived safeness &    &    &    &    &    &    &    &    &    &    &    &    &    &    &  1 &    &    \\
System usability & overall usability &    &    &    &    &    &    &    &    &  1 &    &    &  1 &    &    &  4 &  2 &  2 \\
                                 & perceived ease of use &  1 &  1 &    &    &    &    &    &    &    &    &  1 &    &    &  1 &    &  1 &    \\
                                 & mental demand &    &    &    &    &    &    &    &    &    &    &    &    &    &  2 &    &  1 &    \\
                                 & physical demand &    &    &    &    &    &    &    &    &    &    &    &    &    &  2 &    &    &    \\
                                 & perceived task completion &    &    &    &    &  1 &    &    &    &    &  1 &    &    &    &  2 &    &    &    \\
                                 & effort &    &    &    &    &    &    &    &    &    &    &    &    &    &  2 &    &    &    \\
                                 & frustation &    &    &    &    &    &    &    &    &    &    &    &    &    &  2 &    &    &    \\
                                 & desire to continue the interaction &    &    &    &    &    &    &    &    &  1 &    &    &    &    &  2 &    &  2 &    \\
                                 & user acceptance of the agent &    &    &    &    &    &    &    &    &  1 &    &    &    &    &  2 &    &  2 &    \\
                                 & system consistency &    &    &    &    &    &    &    &    &    &    &    &    &    &    &    &  1 &    \\
                                 & perceived usefulness/helpfulness &    &    &    &    &    &    &    &    &    &    &    &    &    &    &  1 &  4 &  1 \\
                                 & willingness to recommend &    &    &    &    &    &    &    &    &    &    &    &    &    &    &    &  1 &    \\
                                 & motivation &    &    &    &    &    &    &    &    &    &    &    &    &    &    &  1 &    &    \\
UX with CA & overall UX &    &    &    &    &    &    &    &    &    &    &    &    &    &  1 &  1 &  2 &    \\
                                 & contrast of user experience &    &    &    &    &    &    &    &    &    &    &    &    &    &    &    &  1 &    \\
                                 & perceived quality of the interaction &    &    &    &    &    &    &    &    &    &    &    &    &    &    &  1 &  1 &    \\
                                 & perceived quality of CA &    &    &    &    &    &    &    &    &    &    &    &    &    &    &  1 &  1 &    \\
                                 & perceived satisfaction &  1 &    &  1 &    &  1 &    &  1 &    &  1 &    &    &    &  1 &  1 &  3 &  3 &    \\
                                 & perceived engagement &    &    &    &    &    &    &    &    &    &    &    &    &    &    &  1 &  3 &  2 \\
                                 & perceived trust &  1 &    &    &    &    &  1 &    &    &    &    &    &    &    &  1 &  1 &  2 &  1 \\
                                 & confidence &    &    &    &    &    &    &    &    &    &    &    &    &    &    &  1 &    &  1 \\
                                 & perceived capability to control &    &    &    &    &    &    &    &    &    &    &    &    &    &    &    &  1 &  1 \\
                                 & aroused emotions &    &    &    &    &  1 &    &    &    &    &    &    &    &    &    &    &  1 &    \\
                                 & serendipity &    &    &    &    &    &    &    &    &    &    &    &    &    &    &    &    &  1 \\
                                 & pleasant surprise &    &    &    &    &    &    &    &    &    &    &    &    &    &    &    &    &  1 \\
                                 & empathy &    &    &    &    &    &    &    &    &    &    &    &    &    &    &  1 &  1 &  1 \\
                                 & self-reflection/self-awareness &    &    &    &    &    &    &    &    &    &    &    &    &    &  1 &    &  1 &  1 \\
                                 & self-disclose &    &    &    &    &    &    &    &    &    &    &    &    &    &  1 &    &  1 &    \\
                                 & anxiety/tension &    &    &    &    &    &    &    &    &    &    &    &    &    &    &    &  1 &    \\
                                 & comfort &    &    &    &    &    &    &    &    &    &    &    &    &    &    &    &  1 &    \\
                                 & social orientation toward CAs &    &    &    &    &    &    &    &    &    &    &    &    &    &    &  1 &    &    \\
                                 & degree of collaboration  &    &    &    &    &    &    &    &    &    &    &    &    &    &    &    &    &  1 \\
                                 & degree of decision-making &    &    &    &    &    &    &    &    &    &    &    &    &    &    &    &  1 &  1 \\
\bottomrule
\end{tabular}
}

    \label{tab:survey_scale}
\end{table*}

\subsubsection{Interview}
For \textit{dyadic} interaction papers, overall, 17 papers evaluated CAs using interview data analysis. Table ~\ref{tab:interview}. provides an overview of what metrics were evaluated in the user studies. 
On the one hand, users were interviewed regarding their general impressions on the CAs, e.g., ~\citep{kim2019comparing, lee2021exploring, prasad2019dara, lee2020hear, cavazos2019jido},  perceived CA characteristics such as capability in handling requests \citep{folstad2019chatbots}, personality~\citep{kim2019comparing, prasad2019dara}, and trust \citep{large2019s}. On the other hand, they interviewed user behaviors towards conversational agents such as efforts used while interacting with the CAs \citep{lee2021exploring}, actual engagements~\citep{lee2021exploring, ruan2019quizbot}, and daily using practices \citep{lee2020hear}. 

For \textit{polyadic} interaction papers, studies that employed interviews focused more on specific issue-relevant metrics. These metrics include perceived capability of the agent on solving a particular issue~\citep{gulz2011extending}, engagement in the conversation~\citep{avula2018searchbots,seering2020takes,narain2020promoting}, willingness to take actions under the persuasion of the CA~\citep{savage2016botivist}, reflections on selves and others~\citep{seering2020takes, narain2020promoting}. Some general metrics include overall user experience and impression\citep{seering2020takes,narain2020promoting,shamekhi2018face}, as well as potential directions for improvement\citep{narain2020promoting,shamekhi2018face}. Social interaction features seem to be essential for \textit{polyadic} papers, especially factors related to promoting team contributions and engagement, e.g.,  ~\citep{avula2018searchbots}. There are many features that have been evaluated in \textit{dyadic} interactions but are also suitable to be evaluated in the \textit{polyadic} context, such as effect on judgement ~\citep{bawa2020multilingual}, and perceived burden, i.e., time, financial, mental, and emotional burden~\citep{park2020can}.
\begin{table*}[!ht]
    \caption{\textit{Dyadic} ACM Papers using Interview Related Metrics for UX Evaluation }
    \centering
    \resizebox{0.7\textwidth}{!}{
\begin{tabular}{llr}
\toprule
\textbf{Category} & \textbf{Metrics (Defined in Appendix \ref{append-2})} &    \textbf{Number of Papers (Between 2005 and 2021)}\\
\midrule
Perceptions & overall impressions and experience &                       5 \\
          & perceived CA's capabilities &                       1 \\
          & perceived CA appearance and self-presentation &                       1 \\
          & perceived personality &                       2 \\
          & perceived naturalness  &                       1 \\
          & reciprocative &                       1 \\
          & perceived benefits of the agent &                       1 \\
          & motivation to use CAs &                       2 \\
          & perceived burden  &                       1 \\
          & the best and worst aspects of using CAs &                       1 \\
          & perceived human-likeness &                       1 \\
          & perceived communicative experiences &                       1 \\
          & perceived enjoyment &                       1 \\
          & trust &                       1 \\
          & perceived effectiveness &                       2 \\
Behaviors & engagement &                       2 \\
          & notice of CA's certain function &                       1 \\
          & efforts in learning the system &                       1 \\
          & effects on self-warenss, self-reflection &                       1 \\
          & effect on judgement &                       1 \\
          & effects on collaborations &                       1 \\
          & daily practices of using the CA &                       1 \\
\bottomrule
\end{tabular}
}

    \label{tab:interview}
\end{table*}

\subsection{Overlooked Issues of \textit{Polyadic} CAs (RQ6)~\label{RQ-issues}}

We also reviewed overlooked issues and additional findings aside from the primary design focuses, which raise intriguing issues that can be missing in design guidelines. These issues are related to appropriateness, privacy, and ethics in the designs of CAs, which warrant deeper discussions about the role of \textit{polyadic}  CAs, their social influence, and their relationship with human users in different group settings.

\subsubsection{\textit{Polyadic}  CAs Need to be "Visible"}

One overlooked issue points to the users' awareness of the \textit{polyadic}  CAs in the group. \citep{avula2018searchbots} presented a searchbot to support collaborative search tasks, and the authors discussed that the collaborative nature of their searchbot posed a new issue regarding awareness of the CA. They suggested that the searchbot should announce itself and remain "visible" to the users throughout the interaction. 
Whether and how to keep users' awareness of \textit{polyadic}  CAs were discussed in diverse contexts, such as the tutor \citep{chaudhuri2009engaging,ai2010exploring}, the assistant \citep{avula2018searchbots,cranshaw2017calendar}, the moderator or facilitator \citep{kim2021moderator, kim2020bot, isbister2000helper, Hohenstei2018AI}, and one of the members in online communities\citep{savage2016botivist,seering2018social}. The findings suggest a direction regarding design decisions to make users aware of CAs as their peers or not.

\subsubsection{\textit{Polyadic}  CAs Need to be "Ignorable"}

There is some discussion of ignorable design in the reviewed papers. In collaborative learning contexts, \citep{tegos2015promoting} mentioned that CAs are sometimes ignored and abused by the group learner. Authors found that students provided hasty answers to the tutor agent and sometimes wanted to pay more attention to the learning questions instead of the agent's facilitation. Similarly, participants perceived a task reminder CA to be invasive or annoying, as they are "too frequent", "not context sensitive", and distracting \citep{toxtli2018understanding}. 

Only one paper mentioned "easy to ignore" as a design suggestion when developing a supporting agent in multi-user contexts \citep{isbister2000helper}, because if CAs are persistent with their intervention, the effects can potentially backfire. Given that there are limited papers discussing ignorable design suggestions, it remains a crucial issue for designers to create CAs that are less intrusive and are able to detect the moment when users do not want their interventions and shut down properly, particularly when the interaction between human users is the major focus.

\subsubsection{\textit{Polyadic} CAs Need to be Accountable}
Another overlooked issue arises in voice activated CAs in interpersonal spaces\citep{luria2020designing}. As CAs are involved in multiple users' interactions in their homes, participants were confused about how an CA would deal with interpersonal conflicts between users in the home without invading privacy\citep{luria2020designing}. Papers also discussed the issue of CA ownerships - who should CAs be accountable to?  
If a CA is the mediator between human users, how much can and should other users consider this mediator? If there are interpersonal conflicts, what is the standpoint of the CA? What will happen if the agent crosses a perceived boundary, and how should we tackle it? Several questions such as these were asked in the reviewed papers~\citep{seering2020takes, luria2020designing, zheng2021pocketbot}.

\subsubsection{Topics Discussed in \textit{Polyadic} and \textit{Dyadic} Works}\label{RQ-themes}

\begin{figure*}[htbp] 
    \centering           
    \includegraphics[width=0.95\textwidth]{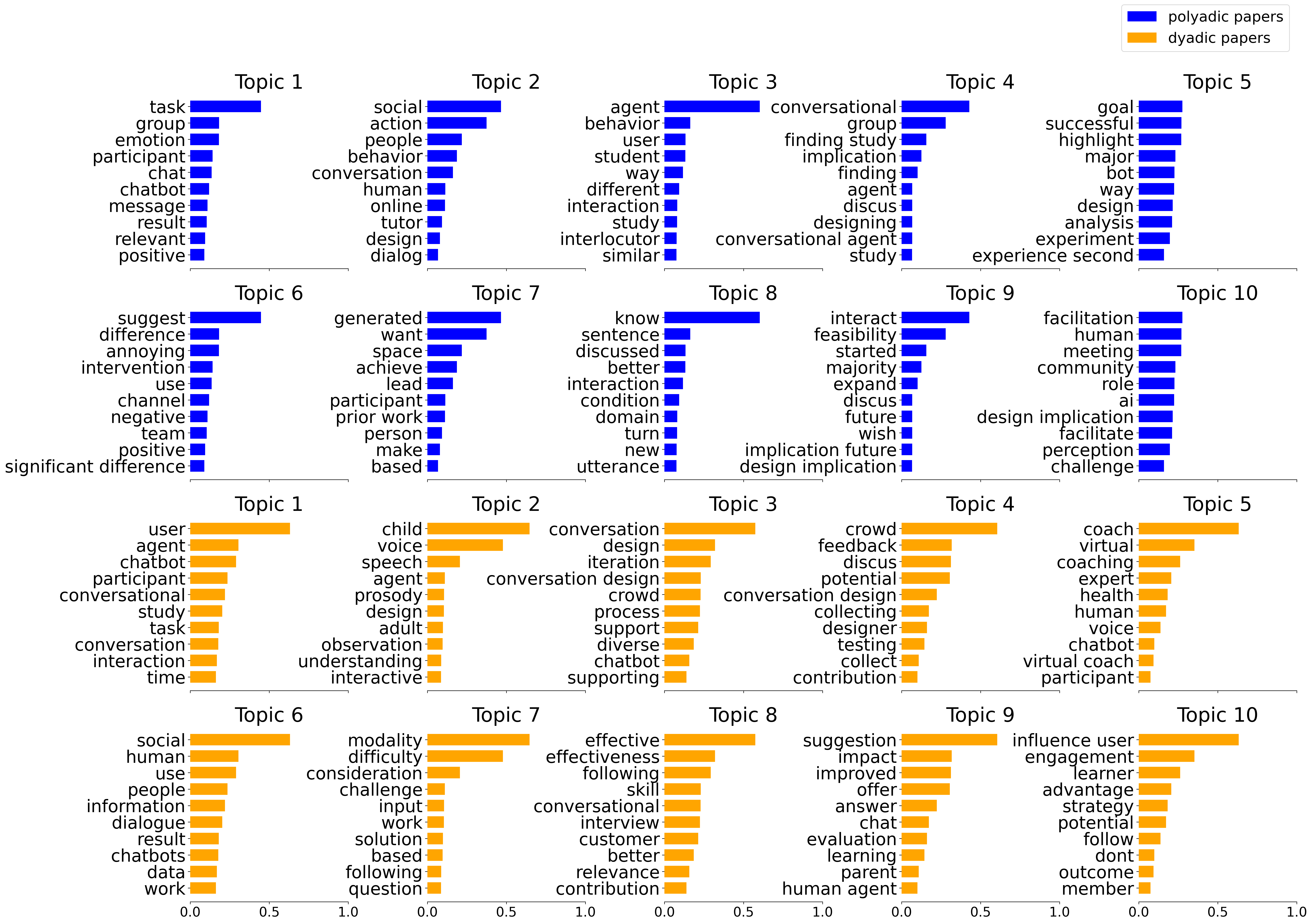}
  \caption{Comparison between topics (ordered by weight) discussed in \textit{polyadic} and \textit{dyadic} papers (the x-axis is the probability distribution of each term in its corresponding topic)}
  \label{topics}
\end{figure*} 

To explore the difference between the topics discussed in the UX research of \textit{dyadic} and \textit{polyadic} CAs, we applied topic modeling method over the papers' \textit{discussion} sections in which authors reflected on the findings and proposed future research directions. As shown in Figure~\ref{topics}, we found that the top topics identified from the  \textit{polyadic} CA studies covered different concepts that were not frequent in the \textit{dyadic} CA articles. 
As shown in Figure~\ref{topics},  \textit{polyadic} papers' Topic 1 was about \textit{group discussion} (18.82\% topic weight); whereas \textit{dyadic} papers mentioned \textit{user-agent interaction} the most (44.16\% topic weight). For example, Peng et al. evaluated a \textit{polyadic} CA \citep{peng2019gremobot} that facilitated emotion regulation in group discussions, which received the highest weight for the \textit{group discussion} topic. This work also contributed to \textit{polyadic} Topic 2 (about \textit{social behavior}, 11.98\% topic weight), as it discussed how to  positively impact users' social behavior and emotion. A similar topic appeared in the \textit{dyadic} works but was ranked much lower in popularity (Topic 6 with  5.86\% weight). On the contrary, Muresan et al.'s paper \citep{muresan2019chats}, a \textit{dyadic} CA weighted the highest for Topic 1 \textit{user-agent interaction}, discussed how anthropomorphism of CAs affects users' engagement. This study also contributed to \textit{dyadic} Topic 2 (\textit{conversational design} with  8.08\% topic weight). However, both topics were not appeared in the top 10 of \textit{polyadic} works. In addition to  \textit{social behavior} (e.g., \citep{rehm2008she, kumar2014triggering}),  \textit{polyadic} CA works tended to discuss more on \textit{education} (Topic 3, 10.94\% weight, e.g., \citep{kim2020bot, dyke2013towards}), and \textit{embodied design} (Topic 5,  10.13\% weight, e.g., \citep{seering2020takes, bohus2010facilitating}). But \textit{education} related topic was also  ranked low in  \textit{dyadic} papers (Topic 10 with 4.49\% weight).

\section{Discussion}
Given the above results of the literature review, we discuss several research opportunities for future design and development of \textit{polyadic} CAs to address social challenges in human-to-human interactions. These opportunities emerge in three main directions, i.e., exploring an under-studied design space, using and developing theoretical foundations, and developing HCI design guidelines for building boundary-aware CAs. We also discuss specific research agendas of future communicative AI technologies~\citep{guzman2020artificial} around two key aspects, e.g., relational dynamics and functional dimensions.

\subsection{\textbf{Spotlight an Under-Explored Design Space}}
Overall, we found a very small fraction of CAs, published in ACM venues, designed for \textit{polyadic} human-AI interaction. Results suggest that such a design space has been severely under-explored. A predominance of CA surveys worked on the classification of CAs~\citep{hussain2019survey,seering2019beyond,laranjo2018conversational,folstad2018different,cassell2000embodied,klopfenstein2017rise,de2020intelligent}, and some efforts were made in investigating CAs' potential to improve user engagement and experience in \textit{dyadic} Human-AI interaction~\citep{chaves2020should, rapp2021human}. However, still little attention was given to the conversational effects of CAs on \textit{polyadic}  human-AI interaction. Thus, this literature review filled the research gap and mapped the progress in this field for future researchers. 

In terms of the application domains, results showed that some areas (e.g., public services, health) fall short of UX research on \textit{polyadic} CAs. Also, user evaluation lacks empirical findings of large scaled human-human interaction. We also found that most original designs were proposed to explore the feasibility of conversational AI in varied application domains (e.g.,~\citep{toxtli2018understanding, dohsaka2009effects,seering2020takes,kim2020bot, benke2020chatbot,luria2020social}. With the tsunami of conversational AI, the need for design knowledge becomes more intense~\citep{rapp2021human}. However, there is no wide consensus on grounded practices on which to create novel designs.

\subsection{\textbf{Connect with and Contribute to Theories}}

Our review highlights a limitation in the theoretical groundings of designing and evaluating CAs for \textit{polyadic} Human-AI interactions. We found that many works did not leverage or contribute to existing theories. This was also found by prior literature reviews of CAs~\citep{rapp2021human,clark2019state}. 
There can be two reasons. First, \textit{polyadic} conversational design is still an emerging area, and HCI researchers have not identified the best theories to explain, build on, and support their studies. Currently, some adopted existing theoretical grounds from social psychology (e.g., CASA)~\citep{rehm2008she}, sociology (e.g., balance theory, structural roles) \citep{nakanishi2003can, seering2018social} and communication studies (e.g., self-extension theory)~\citep{luria2020designing} . Second, \textit{polyadic} CAs are bringing brand-new dynamics of social interactions, to an extent that existing theoretical frameworks cannot account for. For example, prior work~\citep{fox2021relationship} identified that human users' expectations and perceptions towards machines are different from those towards human partners. Thus, when new social dynamics are formed, existing theories may not be sufficient.

To leverage existing theories, scholars need to identify the social interaction \textit{challenges} (e.g., findings in RQ1) to be addressed and the \textit{application domains} (e.g., findings in RQ3), and then look for related theories resolving the particular challenge in the relevant application domain to inspire the CA designs. For example, some \textit{polyadic}  CAs are designed for group members to reach consensus~\citep{kim2021moderator, kim2020bot,benke2020chatbot}. 
Prior works on collaborative engineering, a domain where designs are created to enable engineers to work effectively with all stakeholders for completing collaborative tasks~\citep{borsato2015collaborative}, have applied consensus building theory (CBT)~\citep{briggs2005toward} in designing for collaboration processes~\citep{kolfschoten2007collaboration, nabukenya2009theory}. The CBT is often applied by those conducting IT requirement negotiations or those conducting risk and control self-assessments. In the situation where decisions cannot be made unilaterally because team members are co-responsible peers, teams need to resort to approaches to build consensus to gain commitment~\citep{kolfschoten2007collaboration}. The CBT could be applied to direct four major steps in reaching a consensus, i.e., articulating a proposal, evaluating willingness to commit, diagnosing causes of conflict, and invoking conflict resolution strategies~\citep{briggs2005toward}. Similarly, when designing a \textit{polyadic} CA under a collaborative negotiation situation, designers may leverage the CBT model to inform designs. Future studies can also adopt a similar approach in finding suitable theories or frameworks.

\subsection{Propose the Notion of Boundary-Awareness}

Only a few \textit{polyadic} papers used existing systems in user evaluation. Regardless of the CAs'  originality, the overlooked negative UX identified by prior research indicates that there is a pressing need for improvements in design strategies and guidelines. 

\subsubsection{\textit{Polyadic} CAs are Unique to Design}
In prior works, human-likeness (e.g., empathy \citep{samrose2020mitigating} and self-disclosure \citep{lee2020hear}) is identified as a critical aspect to improving UX and to encouraging users to show favorable feelings (trust, openness, tolerance, etc) towards CAs~\citep{rapp2021human,chaves2020should}. However, similar topics are less reflected by \textit{polyadic}  human-AI interaction (RQ5). Instead, researchers raised most open concerns on how  CAs should establish social boundaries in human-human interactions, which suggests that \textit{polyadic} CAs may be \textit{unique} to design compared with \textit{dyadic} CAs.

Taking the learning context as an example, when a \textit{polyadic} CA was deployed as a study peer, researchers found that students teams often ignore or provide hasty replies to CAs \citep{tegos2015promoting}. However, how can CAs be less intrusive partners in such a context? Similarly, in a family context, when family members have different opinions, how can the CA understand when to "knock the door" when it is tackling conflicts between couples~\citep{zheng2021pocketbot}? Should it let parents know where their children are up to~\citep{luria2020social}? Our findings highlight the importance of questions for CAs to understand social boundaries. Current studies are yet able to take care of the dilemmas in situations where user needs or human-AI interaction conflict and the CAs need to react or take sides.

\subsubsection{Design CAs with Boundary-Awareness} By reviewing the overlooked issues, we identified multi-dimensional themes in \textit{polyadic} human-AI interaction, which we consider as social boundary issues (RQ6). Boundaries in social sciences can be understood as a set of rules followed by most people in a particular society, which are vital in the society because they can guide human behaviors and assist in managing what is and what is not acceptable \citep{houghton2010privacy,lamont2002study}. This issue is closely related to privacy and disclosure, as the presence of \textit{polyadic} CAs as social actors changes the social dynamics and users' information management strategies.

Communication privacy management theory (CPM)~\citep{petronio2013brief} identified three main elements in people's private information management: (1) privacy ownership, (2) privacy control, and (3) privacy turbulence. When an individual has decided to disclose private information, as a result, the recipient of the information becomes a co-owner or shareholder. From that moment, the initial owner of the information must set rules and boundaries of how to manage private information. Privacy turbulence may happen when these rules are violated, and the original owner may refrain from disclosing any more information or may engage in negotiations or coordinate rules and boundaries with the co-owner~\citep{petronio2014communication}. 
Clarifying ownership of users' private data to \textit{polyadic} CAs and enabling them to learn privacy boundaries in relationships over time are essential steps towards building social sensibility \textit{polyadic} CAs when participating in human-human interactions. For example, a boundary-aware CA can learn if and when it should intervene when mediating couples' conflicts by receiving requests from one side of the couple~\citep{zheng2021pocketbot}. Similarly, when users request a CA to understand situations of their living alone elderly parents, the boundary aware CA can reply appropriately. The privacy boundaries in the interactions also depend on how CAs present themselves and how users perceive the CAs. For instance, when CAs join human interactions, their roles may vary from active conversation participants to less salient group collaboration assistants. A boundary-aware CA may adjust its proactivity and intensity in joining the discussion.

In our review, multiples studies showed that during the interactions, users expressed confusions and concerns about what roles \textit{polyadic} CAs should play, how \textit{polyadic} CAs should behave and what is the proper distance that \textit{polyadic} CAs should maintain with their users in various group settings. As \textit{polyadic} human-AI interaction involves multiple relationships and complex social dynamics in its nature \citep{tegos2015promoting}, the boundary between CAs and different users, and the boundary between users need to be taken care of.

Thus, we propose to design CAs with \textit{boundary-awareness} for supporting \textit{polyadic}  human-AI interaction. Traditional HCI design addressed boundaries, e.g., between computers and people and between computers and the physical world \citep{vogel2004interactive,stephanidis2019seven}. Different from the traditional boundaries, which are often between virtual and physical worlds \citep{burden2016conceptualising}, the new boundaries brought by the CAs involve human-to-human boundaries. Prior work~\citep{mou2017media} suggested that, when facing AI, humans demonstrate different personalities from interacting with other humans. 
Thus, it might be difficult for CAs to add boundary management in the design because of the collective result of unclear roles, undecided social rules, and social emotions \citep{guzman2020ontological} during human-AI interaction.

CAs with \textit{boundary-awareness} cannot be detached from the discussion of privacy, e.g.,~\citep{avula2018searchbots, luria2020social, reig2020not}. To address the boundary issue such as "whether the CA should inform the parents where their kids are up to" \citep{luria2020social}, we can gain insights from privacy boundaries. 
For example, Palen and Dourish~\citep{palen2003unpacking} discussed three boundaries with different desires during one's privacy management. 1) "Disclosure boundary" is the desire to keep one's information private or public. Privacy regulation in practice is not simply a matter of prohibiting one's data from being disclosed. Instead, human-to-human interactions frequently require selective disclosures of personal information to declare allegiance or even differentiate ourselves from others. 2) "Temporal boundary" suggested that privacy management is in the tension between past, present, and future. Users' response to disclosure situations is interpreted according to other events and expectations. 3) "Identity boundary" implied a boundary between self and other. For example, employees are discouraged from using corporate email addresses to post to public forums. Designers can leverage these three aspects to understand the most critical privacy boundaries in creating CAs with boundary awareness.

\subsubsection{Adopt and Evaluate the Existing Guidelines}
The unique challenges urge us to re-examine existing AI design guidelines and design notions. For example, Microsoft's guidelines for designing human-AI interaction suggested that AI-infused systems should "support efficient dismissal"~\citep{amershi2019guidelines}, therefore CAs should be able to learn the social boundaries such that their interaction is perceived less intrusive, addressing a concern that was raised in prior \textit{polydic} CA works~\citep{tegos2015promoting, beuth2010software}. The guidelines also suggested that AI systems should "match relevant social norms" (i.e., "the experience is delivered in a way that users would expect") and should "mitigate social biases" (i.e., "the system language and behaviors do not reinforce undesirable and unfair stereotypes and biases given their social and cultural context") ~\citep{amershi2019guidelines}.  An example of potentially using these guidelines can be found in prior \textit{polydic} CA works~\citep{kim2021moderator} which designed a CA tailored for structural group discussions. The study reported that the CA facilitates the team reaching a logical consensus on a highly contentious topic even though the members'  positions and understandings are vehemently opposed to each other, indicating the design of CA's that acts in a way that the team expected (knowing the norm).  However, the authors also proposed that when discussing divisive issues, it may be more appropriate to design with interpersonal and social power dynamics in mind and encourage and protect participants' contributions from marginalized groups (understanding social biases.)

Moreover, existing CA guidelines also provide a broad set of items for designing responsible and trustworthy CAs, such as transparency, human control, awareness of human values, accountability, fairness, privacy and security, accessibility, and professional responsibility, e.g.,~\citep{microsoft2018}. These guidelines can be employed in future CA designs, and more academic works should also evaluate these guidelines'  effectiveness in the wild.

\subsection{Other Key Aspects of Communicative AI} 
We also discuss research agendas of future communicative AI technologies from two perspectives, i.e., relational dynamics and functional dimensions. 

\subsubsection{Relational Dynamics}
Findings from RQ4 suggest that \textit{polyadic} CAs show potentials for influencing social dynamics. Relational dynamics reflect the ways people interact with communicative technologies, with themselves, and with other people or group of people~\citep{guzman2020artificial}. Possible research directions include the dynamics of relationship types, and relational attributions. We suggest that \textit{polyadic} CAs can be further investigated in its effectiveness to foster new relational dynamics. For example, according to the Computers as Social Actors paradigm~\citep{nass1994computers}, CAs are considered as "social actors". Humans rely on many perceivable attributes during an social interactions with other humans \citep{donath2007signals}. To make sense of human-AI interaction, prior works identify, implement, and test CA effects in various social attributions, such as "social cue"~\citep{ochs2017you, feine2019taxonomy} (i.e., a signal that triggers a social reaction of the user towards the CA), "social roles"~\citep{seering2018social} (i.e., the function CAs and humans play in the interaction context), and social identities~\citep{mirbabaie2021understanding, wambsganssethical} (i.e., how CA or humans organize themselves into and within groups, social values that take collective goals, ethical concerns into consecrations), to name a few.

\subsubsection{Functional Dimensions} 
We propose that the design and use of \textit{polyadic} CAs should address questions such as: 1) what communication challenges are to be addressed by \textit{polyadic} CAs?; 2) what is the unique function of \textit{polyadic} CAs?; and 3) how  can \textit{polyadic} CA effectively address the identified challenge compared to other methods? These questions should fit into the functional dimensions proposed by~\citep{guzman2020artificial}, which reflects how certain AI technologies are designed and how people can make sense of these devices and applications.

In our literature review, not all works identified a fundamental challenge of human-human interaction before proposing a \textit{polyadic} CA (RQ1). 
There could be a tendency of technology-determinism", believing that "technological change can determine social change in a prescribed manner\citep{dafoe2015technological}." Namely, some works assumed that CAs could help with the challenges without evaluating the effect of the proposed CAs by comparing with other counterparts, e.g., without a CA, with a human, or with previously tested CAs.  In future research, designers and practitioners may be better aware of this potential tendency and propose more comprehensive study plans to evaluate the proposed CAs.

\section{Limitations and Future Work}

This literature review has several limitations as a result of the data source, the search query, and data analysis methods. First, we chose ACM DL as the source of our literature review work. Several prior works also conducted literature reviews by using only ACM publications for HCI and CSCW ~\citep{wainer2007empirical, pinelle2000review} and computer science research ~\citep{wainer2009empirical}. However, using one major source could potentially generate a selection bias, e.g., including certain publications from non-ACM venues into the analysis could be challenged as "distorting the conclusions toward a particular direction"~\citep{wainer2009empirical}. Therefore, when presenting our findings, we focused on the selected papers as samples illustrating the emerging themes that were not presented in prior literature reviews. One major purpose of doing so is to inform UX researchers and practitioners to design conversational AI by leveraging the existing UX research outcomes and apply the suite of measurements in their work. In the future, conducting a systematic literature review by sampling from different sets of HCI research, e.g., papers indexed by IEEE, Web of Science, Scopus, and Science Direct, can further deepen our understanding of conversational AI research.

Second, we selected research publications that completed both design and UX evaluation of CAs. According to quantitative survey of experimental evaluation in computer science,  scholarly publications in ACM could be classified into five categories, including formal theory, design and modeling, empirical work, hypothesis testing, and others (e.g., surveys)~\citep{tichy1995experimental}. The publications we selected could be a combination of design and modeling~(i.e., "systems, techniques, or models, whose claimed properties cannot be proven formally") and empirical work (i.e., "articles that collect, analyze, and interpret observations about known designs, systems, or models, or about abstract theories or subjects~\citep{wainer2007empirical}"). Even though we tuned the search query to the best we could during web-crawling, it is possible that certain works were not captured by the terms we defined due to our fields of expertise. Meanwhile, it is expected that a significant amount of works that proposed novel techniques without UX evaluations were not included in our analysis. Therefore, we do not claim that the findings are exhaustive. Even though our findings build upon significant prior literature review results, the themes of \textit{polyadic} challenges and evaluation metrics can very and will expand with more works identified.  

Lastly, the thematic analysis process is subjective, and lemmatization \citep{nguyen2015topic} and topic modeling \citep{gurcan2021mapping}  might introduce potential biases as well. For example, lemmatization reduces words to their lemma forms, the use of lemmatization could potentially cause a loss of words' meaning at a certain level. For topic modeling, since we used text from the original papers without filtering irrelevant words, the results contained noises and irrelevant words that sometimes interfered the interpretability. Therefore, we cannot claim differences to be statistically significant. For example, the topic modeling results were discussed in the context of specific examples to elicit more discussions and future research. 

\section{Conclusion}

This literature review presents what fundamental human-human interaction challenges are addressed by \textit{polyadic} CA works scrutinized from ACM publications, what effects on human-human interaction are evaluated by these works, and what issues are overlooked when designing  \textit{polyadic} CAs. In particular, we propose that researchers and practitioners should design CAs with \textit{boundary awareness} for supporting \textit{polyadic} human-AI interaction. Further, we envision several research opportunities on conversational human-AI interaction with respect to the theoretical groundings, relational dynamics, and functional dimensions.

\section{Acknowledgement}

This material is based upon work supported by the National Science Foundation under Grant No. 2119589. The research is also partially supported by the IBM-ILLINOIS Center for Cognitive Computing Systems Research (C3SR), a research collaboration as part of the IBM AI Horizons Network.

\clearpage
\appendix
\onecolumn
\section{APPENDIX}
\label{append}

\subsection{Definitions of Different Metrics Used in \textit{Polyadic} Papers}
\label{append-1}

\begin{table}[!ht]\centering
\label{tab:metric_def_polyadic}
\resizebox{\textwidth}{!}{
\begin{tabular}{p{3cm}|p{6cm}|p{13cm}}\toprule
\textbf{Category} &\textbf{Metrics} &\textbf{Definition} \\\midrule
Sociability & discussion quality & the quality of users' discussions \\\cmidrule(lr){2-3} 
&consensus reaching & the reaching of a consensus in terms of behavioral and perceived opinion alignment \\\cmidrule(lr){2-3} 
&opinion expression & users' message quantity, even participation, and perceived outspokenness \\\cmidrule(lr){2-3} 
&opinion diversity & the number of unique lexical morphemes shared within a group \\\cmidrule(lr){2-3} 
&even participation & how equally individual members contribute to the discussion. \\\cmidrule(lr){2-3} 
&linguistic features & the linguistic features extracted from users' utterance text, e.g. verbosity, readability, sentiment, linguistic diversity and adaptability \\\midrule 
Task-related &task completion time & the time elapsed before users successfully completed the intended task(s).\\\cmidrule(lr){2-3}
&efficiency & the number of system turns required and average system reaction time to users' requests. \\\cmidrule(lr){2-3}
& effectiveness & users' successes in the tasks undertaken with the CA support.\\\cmidrule(lr){2-3} 
&satisfaction & users' overall ratings for CA's understandability, relevancy and  efficiency\\\cmidrule(lr){2-3} 
&team performance  & the team's success in the tasks undertaken with the CA's support \\\midrule
Communication quality &perceived communication effectiveness & the degree of how much the CA helps users reach the goal \\\cmidrule(lr){2-3} 
&perceived communication fairness & the degree of participation fairness that users perceive during discussions \\\cmidrule(lr){2-3} 
&perceived communication efficiency & how quickly the consensus is reached during discussions \\\cmidrule(lr){2-3} 
&perceived group climate & the relatively enduring tone and quality of group interaction experienced similarly by group members \\
\midrule
Socio-emotional &perceptions of other group members & users' opinion and perception about other members in group discussions\\\cmidrule(lr){2-3} 
&perception towards the agent& users' opinion and perceptions about the CA in discussion \\\cmidrule(lr){2-3} 
&perceptions about the collaboration & users‘ collaboration experience in terms of users' awareness of each other's activities, effort, and enjoyment. \\\cmidrule(lr){2-3} 
&perceived social presence & users' experience of being present with other persons and having access to their thoughts and emotions \\\cmidrule(lr){2-3} 
&perceived social support & users' experience of being provided with support from other persons during discussion\\\cmidrule(lr){2-3} 
&level of annoyance with the intervention &  users' perceived annoyance when receiving the CA's intervention\\\cmidrule(lr){2-3} 
&opinion of oneself/partner/cultural stereotypes & users' opinion about self, partner, and cultural stereotypes under the CA's influence during discussion\\\cmidrule(lr){2-3} 
&perceived appropriateness of CA's social behavior & users' perception about the CA's ability to behave appropriately like a human\\\cmidrule(lr){2-3} 
& psychological wellbeing & users'  psychological wellbeing measured by positive relationships with others, personal mastery, autonomy, a feeling of purpose and meaning in life, and personal growth and development\\\midrule

Traditional&anthropomorphism& the extent to which the CA can demonstrate attribution of human characteristics or behavior \\\cmidrule(lr){2-3} 
&perceived emotional competence & users' perception of the CA's emotional skills\\\cmidrule(lr){2-3} 
&unmet expectations & users' expectations that are not met by the CA during the study \\\cmidrule(lr){2-3} 
&privacy concerns & users' concerns about the safeguarding and usage of personal data provided to the CA \\\cmidrule(lr){2-3} 
&level of perceived conversation control & users' feeling in control of the discussion using the CA\\\cmidrule(lr){2-3} 
&perceived satisfaction & users' overall ratings for CA's understandability, relevancy and efficiency \\\cmidrule(lr){2-3} 
&performance/effectiveness& the measure of user's success in the tasks undertaken in the CA interactions \\\cmidrule(lr){2-3} 
&level of engagement& users' level of engagement with the CA, e.g., the number of interactions measured by clicks and selections \\\midrule
Sociability&perceived capability to promote contributions& users' perception about the CA's ability to elicit contributions and opinions from participants during discussion\\\midrule
Issue-relevant&decisions to (not) engage with the CA& users' decision of whether users would like to engage in interactions with the CA\\\cmidrule(lr){2-3} 
&overall UX& users' perception of the overall user experience during interactions with the CA\\\cmidrule(lr){2-3} 
&reflections on selves and others& users' reflections on behaviors and feelings of self or other participants\\\cmidrule(lr){2-3} 
&willingness to take actions& users' willingness to take certain actions under the persuasion of the CA\\\cmidrule(lr){2-3} 
&direction of improvement& the possible directions of improvements proposed by users' after interacting with the CA\\

\bottomrule
\end{tabular}
}
\end{table}

\clearpage
\subsection{Definitions of Different Metrics Used in \textit{Dyadic} Papers}
\label{append-2}
\begin{table}[!ht]
\small\centering
\label{tab:metric_def_dyadic}
\resizebox{\textwidth}{!}{
\begin{tabular}{p{3cm}|p{4cm}|p{10cm}}\toprule
\textbf{Category} &\textbf{Metrics} &\textbf{Definition} \\\midrule
Linguistic features &positive/negative words & number of users' use of positive/negative words to express emotions \\\cmidrule(lr){2-3} 
&length of utterances & users' number of users' use of words per statement \\\cmidrule(lr){2-3} 
&personal pronouns & users' use of first- and third-person pronouns \\\cmidrule(lr){2-3} 
&term-level repetition & the proportion of terms in one utterance which were repeated from the participant’s previous utterance \\\cmidrule(lr){2-3} 
&utterance-level repetition & the proportion of utterances where term-level repetition is greater than zero\\\cmidrule(lr){2-3} 
&use of concrete words & the concreteness of each entry, e.g., "tennis" is more concrete than "sports" \\\cmidrule(lr){2-3} 
&variation in language used & the variation in expressions and use of different words\\\midrule
Prosodic features &variation in speech-based agents & the word count, variance in pitch, rate and loudness in audio interactions\\\midrule
Dialogue features &dialogue/response quality & the syntactic readability and intensity of sentiments in users' replies\\\cmidrule(lr){2-3} 
&dialogue efficiency metrics & the number of system turns required and average system reaction time to users' requests\\\cmidrule(lr){2-3} 
&expressiveness & the quality of effectively conveying thoughts or feelings by users \\\midrule
Tasks related &task fulfillment rate/effectiveness & the measure of users' success rate in tasks undertaken during the interactions \\\cmidrule(lr){2-3} 
&system use & a mixed measure of multiple system-related metrics, e.g., types of messages, words, time taken to complete the task\\\cmidrule(lr){2-3} 
&user entry time & the amount of time users need to provide inputs to or interact verbally with the CA \\\cmidrule(lr){2-3} 
&dropout rate& the percentage of respondents who quit before the study was completed \\
\midrule
Algorithm &intent modeling evaluation & the accuracy of which the CA can identify the correct intents from users' utterances \\\cmidrule(lr){2-3} 
&error rate & the rate of errors occurred during users' interaction with the CA \\\midrule
User &self-disclosure & users' quality of responses to the CA based on their trust towards the CA \\\cmidrule(lr){2-3} 

&intimacy & users' perceive closeness,inter-connectedness, and companionship from the CA\\\cmidrule(lr){2-3} 
&reflection & users' self-rated frequency of reflecting on thoughts and consciousness of their inner feelings\\\cmidrule(lr){2-3} 

&impolite/aggressive behaviors & the number of occurrences of impolite phrases\\\cmidrule(lr){2-3} 
&motivation & users' motivations or intents that drive users to interact with the CA \\\cmidrule(lr){2-3} 
&affective states & the proportion of time users spent in each affective state\\

\bottomrule
\end{tabular}}
\end{table}

\begin{table*}[!ht]\centering
\resizebox{\textwidth}{!}{
\begin{tabular}{p{3cm}|p{5.5cm}|p{10cm}}\toprule
\textbf{Category} &\textbf{Metrics} &\textbf{Definition} \\\midrule
Conversation related &syntactic readability &the syntax-wise readability of conversation text between users and CA based on the Flesch-readability score \\\cmidrule(lr){2-3} 
&self-reported response quality & the perceived response quality reported by users \\\cmidrule(lr){2-3} 
&informativeness/information quality & the quality of information conveyed by the CA through text \\\cmidrule(lr){2-3} 
&conversation smoothness & the level of smoothness of the conversation between users and CAs measure by Session Evaluation Questionnaire\\\cmidrule(lr){2-3} 
&intensity of sentiments & the intensity of sentiments expressed by users calculated by TextBlob \\\cmidrule(lr){2-3} 
&instruction quality & the quality of instructions given by the CA to users  \\\midrule

Perception towards the CA &perceived common ground & the perceived mutual understanding between users and the CA \\\cmidrule(lr){2-3} 
&opinion of the CA as a partner & the ease of which subjects were able to interact with the CA \\\cmidrule(lr){2-3} 
&playfulness/enjoyment & the extent to which the CA can match users' interests \\\cmidrule(lr){2-3} 
&sociability & users' perception of the CA's social skills \\\cmidrule(lr){2-3} 
&perceived anthropomorphism& the extent to which the CA can demonstrate attribution of human characteristics or behaviors \\\cmidrule(lr){2-3} 
&perceived warmth of the AI system & the extent to which users feel the CA is good-natured and warm \\\cmidrule(lr){2-3} 
&impression & users' feelings of the CA regarding its competence, confidence, warmth, and sincerity \\\cmidrule(lr){2-3} 
&perceived personality traits & users' feelings of the CA in terms of openness, conscientiousness, extraversion, agreeableness, and neuroticism\\\cmidrule(lr){2-3} 
&perceived persuasiveness & users' perception of the CA's utterances rated as bad–good, foolish–wise, negative–positive, beneficial–harmful, effective–ineffective, and convincing–unconvincing\\\cmidrule(lr){2-3} 
&perceived intimacy & users' perceived intimacy and closeness with the CA, e.g., feeling of inter-connectedness of self and other\\\cmidrule(lr){2-3} 
&perceived naturalness &  users' agreement level to the statement that the CA's responses are natural \\\cmidrule(lr){2-3} 
&perceived safeness & users' sense of safety when interacting with the CA \\\midrule

System usability&overall usability&users' perceived overall usability of the CA \\\cmidrule(lr){2-3} 
&perceived ease of use & the degree to which users believe that interacting with the CA would be free of effort \\\cmidrule(lr){2-3} 
&mental demand & the amount of mental or perceptual activities (e.g., thinking, deciding, calculating, remembering, looking, searching) that is required to interact with the CA
\\\cmidrule(lr){2-3} 
&physical demand & the amount of physical activities (e.g., pushing, pulling, controlling, activating, etc.) that is required to interact with the CA  \\\cmidrule(lr){2-3} 
&perceived task completion & the degree to which users believe to have successfully communicated and reached a mutual understanding with the CA \\\cmidrule(lr){2-3} 
&effort & the total workload associated with the tasks, considering all sources and components \\\cmidrule(lr){2-3} 
&frustration & users' feelings of being insecure, discouraged, irritated, and annoyed versus being secure, gratified, content and complacent when interacting with the CA\\\cmidrule(lr){2-3} 
&desire to continue the interaction & the degree to which users would consider keep using this method in the future, or users' behavioural tendencies through their desire to help and cooperate with the CA \\\cmidrule(lr){2-3} 
&user acceptance of the agent &  users' willingness to accept the CA's interaction \\\cmidrule(lr){2-3} 
&system consistency & the consistency between the behaviors and utterances of the CA\\\cmidrule(lr){2-3} 
&perceived usefulness / helpfulness & users' own perceptions of the session’s efficacy, e.g., the CA gave users good suggestions for helping them discover songs.\\\cmidrule(lr){2-3} 
&willingness to recommend & the degree to which users would recommend the CA to their friends or family for managing mental well-being and to people who have needs\\\cmidrule(lr){2-3} 
&motivation & users' motivation or intent to interact with the CA\\
\bottomrule
\end{tabular}}
\end{table*}

\begin{table*}[!ht]\centering
\resizebox{\textwidth}{!}{
\begin{tabular}{p{3cm}|p{6cm}|p{12cm}}\toprule
\textbf{Category} &\textbf{Metrics} &\textbf{Definition} \\\midrule
User experience with the CA &user experience (UX) & a mix ratings of the general experience, e.g., emotion, ease of use, usefulness, and intention to use\\\cmidrule(lr){2-3} 
&contrast of user experience & users' perceptions of the experience without drawing explicit attention to the contrast between their expectations and their experience \\\cmidrule(lr){2-3} 
&perceived quality of the interaction & users' overall self-rated quality of the CA, e.g., in communicating, building rapport, and task fulfillment's \\\cmidrule(lr){2-3} 
&perceived satisfaction & users' overall ratings for CA's understandability, relevancy and efficiency\\\cmidrule(lr){2-3} 
&perceived engagement & the degree that the CA can engage the participants during the conversation, e.g., making users feel it is entertaining and interesting to engage in a dialogue with the CA \\\cmidrule(lr){2-3} 
&perceived trust & the CA's ability in providing unbiased and accurate suggestions and making users trust it  \\\cmidrule(lr){2-3} 
&confidence & users' confidence that users will like the content the CA suggests \\\cmidrule(lr){2-3} 
&perceived capability to control & users' feeling of being in control of the conversation \\\cmidrule(lr){2-3} 
&aroused emotions & change of users' emotions state while using the system \\\cmidrule(lr){2-3} 
&serendipity & the CA's ability in recommending things that users had not considered in the first place but turned out to be a positive and surprising discovery\\\cmidrule(lr){2-3} 
&pleasant surprise & the CA's ability in providing contents that are overall pleasantly surprising to users \\\cmidrule(lr){2-3} 
&empathy & the CA's ability to understand and share the feelings of users
\\\cmidrule(lr){2-3} 
&self-reflection/self-awareness & users' reflection on thoughts and consciousness of their inner feelings \\\cmidrule(lr){2-3} 
&self-disclose & the degree of which users feel comfortable about disclosing to the agent and express opinions openly
\\\cmidrule(lr){2-3} 
&anxiety/tension & the degree of which the interaction makes users feel anxious or tense \\\cmidrule(lr){2-3} 
&comfort & the degree of which the interaction is comfortable \\\cmidrule(lr){2-3} 
&social orientation toward CAs & the desire to engage in human-like social interactions with CA, which is associated with a mental model of an agent system as being a sociable entity \\\cmidrule(lr){2-3} 
&degree of collaboration & the level of collaboration between users and the CA during interaction\\\cmidrule(lr){2-3} 
&degree of decision-making & the degree of which the CA helps with users' decision-making\\\midrule

Perceptions &overall impressions and experience& users' perceptions of the overall experience interacting with the CA \\\cmidrule(lr){2-3} 
&perceived CA's capabilities & the CA's capability in handling simple requests and resembling human representative \\\cmidrule(lr){2-3} 
&perceived CA appearance and self-presentation & the CA's visual appearance and persona \\\cmidrule(lr){2-3} 
&perceived personality & users' feeling for the CA in openness (intellectual curiosity, creativity), conscientiousness (neatness, perseverance, reliability, and responsibility), extraversion (sociability, activity, and assertiveness), agreeableness (friendliness, helpfulness, and cooperativeness in dealing with others) and neuroticism (stability, anxiety, and the frequency of experiencing negative affect)\\\cmidrule(lr){2-3} 
&perceived naturalness & the degree of which users feel the conversation with the CA is natural, not forced\\\cmidrule(lr){2-3} 
&reciprocative & the degree of which users feel the CA reciprocated their language or feelings \\\cmidrule(lr){2-3} 
&perceived burden & the degree of which users feel the conversation with the CA is costly in time, financially, mentally and emotionally. \\\cmidrule(lr){2-3} 
&perceived human-likeness & the CA's ability to talk like a human and its conversational skills \\\cmidrule(lr){2-3} 
&perceived effectiveness & the degree of how much the CA helps to address users’ needs \\\midrule
Behaviors & engagement & the degree that the CA can engage the participants during the conversation, e.g., measured by an engagement rating between 1 (not engaging) and 5 (very engaging) in user survey \\\cmidrule(lr){2-3} 
&notice of CA's certain function & the action of the CA sending messages informing what the chatbot could do, noticing a tutorial and menu\\\cmidrule(lr){2-3} 
&efforts in learning the system & efforts users take in learning how to interact with the CA\\\cmidrule(lr){2-3} 
&self-reflection/self-awareness & users' reflections on thoughts and consciousness of their inner feelings\\\cmidrule(lr){2-3} 
&effects on judgement & the degree that users feel the CA affected their evaluation positively, negatively or neither\\\cmidrule(lr){2-3} 
&effects on collaborations & the degree of which the CA affected users' willingness of collaborating with the CA\\\cmidrule(lr){2-3} 
&daily practices of using the CA & participants’ daily practices of using the chatbot, e.g., "Please
briefly tell us how you used this chatbot during the past three
weeks"\\
\bottomrule
\end{tabular}}
\end{table*}